\documentclass{article}
\usepackage{arxiv}
\usepackage{authblk}
\usepackage{natbib}
\usepackage{url}
\usepackage{hyperref}
\usepackage{paralist}
\usepackage{marginnote}
\usepackage{mathtools}
\usepackage{amssymb}
\usepackage{svg}

\title{Coherent Source Subsampling: A Data-Driven Strategy for Restoring Causal--Acausal Symmetry in Ambient Seismic Wavefield Correlations}

\author[1]{Sanket Narayan Bajad}
\author[2]{Pushkar Bharadwaj}
\author[3]{Pawan Bharadwaj}

\affil[1]{
Centre for Earth Sciences, Indian Institute of Science, 

Bengaluru, Karnataka 560012, India

\texttt{sanketbajad@iisc.ac.in}
\vspace{1em}
}
\affil[2]{
Centre for Earth Sciences, Indian Institute of Science, 

Bengaluru, Karnataka 560012, India

\texttt{pushkarbharadwaj537@gmail.com}
\vspace{1em}
}
\affil[3]{
Centre for Earth Sciences, Indian Institute of Science, 

Bengaluru, Karnataka 560012, India

\texttt{pawan@iisc.ac.in}
}

\begin{document}
\maketitle

%
%

\begin{abstract}

Ambient noise tomography relies on the assumption that the seismic wavefield is equipartitioned. In practice, ambient noise sources are spatially and temporally heterogeneous, producing biased estimates of Green's function between stations. We introduce a data-driven method, Coherent Source Subsampling (CSS), which selects and averages only cross-correlation time windows associated with excitation of sources in the stationary zone. By restricting the ensemble average to these windows, CSS mitigates effects of non-uniform source distribution and restores causal–acausal symmetry in the retrieved interstation response. Applications to regional ambient-noise datasets show that CSS stabilizes surface-wave dispersion measurements even when source statistics violate assumptions of standard seismic interferometry. For the central California dataset, CSS-derived group-velocity tomograms consistently image a high-velocity block between the Rinconada and San Andreas faults across multiple periods. In comparison, the full-ensemble (linear) average does not capture this block, which is well established. Our approach is particularly useful for short-duration passive surveys.
\end{abstract}

\section{Introduction}

Surface-wave tomography has long been a
cornerstone of the imaging of the Earth's crust and upper mantle.
The resolution of traditional surface-wave tomography, which is based on seismic waves generated by earthquakes recorded by regional and global seismic networks, is fundamentally constrained by the spatial and temporal distribution of seismicity~\citep{trampert2003global,ritzwoller1998eurasian,ritzwoller2001crustal,shapiro2002monte,ekstrom2011global, mordret2013near}.
Over the last 20 years, ambient seismic noise cross-correlations have supplemented surface wave tomography~\citep{shapiro2004emergence,sabra2005surface,yang2007ambient,lin2008surface,yang2008structure,yao2009analysis,ritzwoller2011ambient}.

Ambient noise tomography requires cross-correlating seismic noise recorded at pairs of stations. The averaged cross-correlations are then typically 
inverted for the subsurface velocity structure between the stations.
In this paper, $\mathbf{w}_{ij}^{(m)}$ denotes the cross-correlation vector computed between stations $i$ and $j$ over a finite time window indexed using $m$:
\begin{equation}
\mathbf{w}_{ij}^{(m)} 
= \mathbf{s}^{(m)} \ast \mathcal{G}_{ij}\!\left(\mathbf{x}_s^{(m)}\right).
\label{eq:w_def}
\end{equation}
Here, $\ast$ denotes the temporal convolution.
We assume a suitable duration of ambient noise recording such that the wavefield within each time window can be approximated as being generated by an effective noise source located at $\mathbf{x}_s^{(m)}$ with $\mathbf{s}^{(m)}$ as its autocorrelation vector.
This assumption is reasonable when the window length is short
compared to the timescale over which the source distribution evolves.
Our analysis shows that 30-minute windows are sufficiently short for crustal-scale ambient-noise applications.
In our notation, the function $\mathcal{G}_i$ outputs the discrete time-domain Green's function associated with an impulsive source at $\mathbf{x}_s^{(m)}$ and evaluated at the $i$th station, and the function $\mathcal{G}_{ij}$ outputs the cross-correlation between the outputs of $\mathcal{G}_i$ and $\mathcal{G}_j$.
For simplicity, our discussion focuses on a single component of the seismic wavefield (e.g., vertical component for Rayleigh waves), though the framework can be readily extended to other components or multi-component analysis. 

\begin{figure}
\noindent\includegraphics[width=\textwidth]{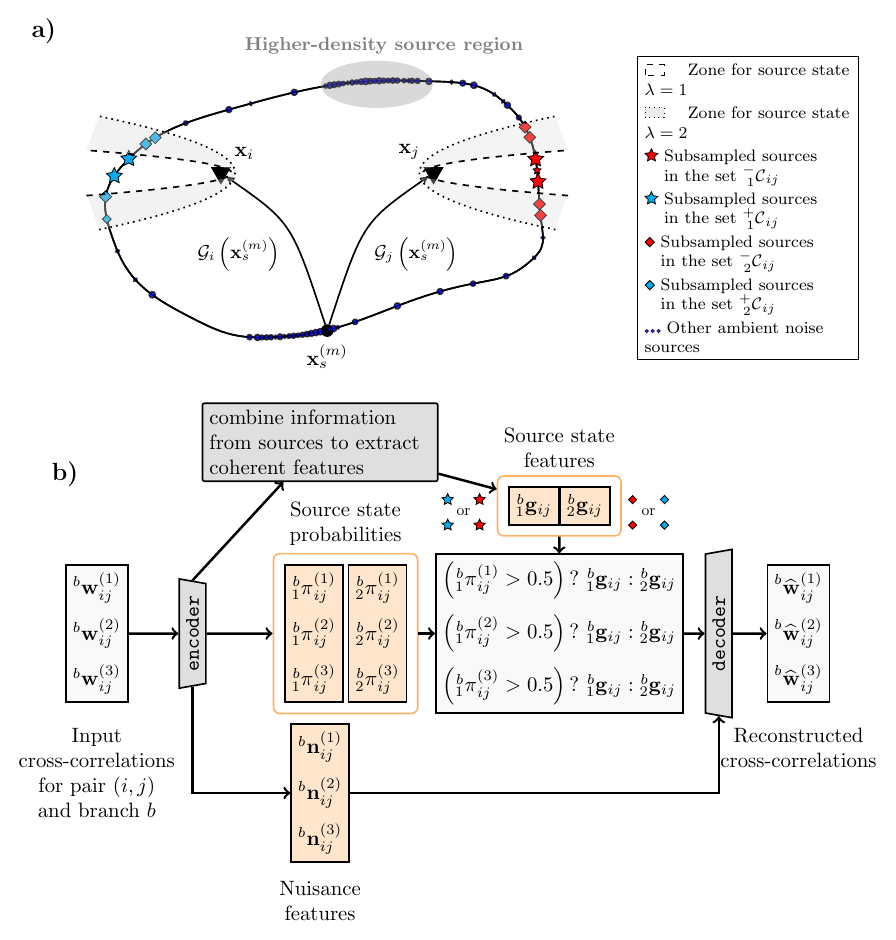}
 \caption{
 Coherent source subsampling (CSS) concept.
       a)  Ambient noise sources on the boundary, distributed non-uniformly, excite receivers at $\mathbf{x}_i$ and $\mathbf{x}_j$. The dashed line marks the source state $\lambda$=$1$ associated with the stationary‑phase zone. For this state, red (acausal) and cyan (causal) stars denote subsampled sources that are considered for averaging. Diamonds and a dotted line represent the second state ($\lambda$=$2$).
       CSS excludes
       the remaining blue-dot sources while averaging as they 
       bias the estimation of interstation response.
       b)
    Latent‑variable autoencoder model for CSS. For each branch $b$ and pair $(i,j)$, the encoder maps input cross-correlations to three latent-variable components (orange boxes). 
    The source-state probabilities act as data‑driven selectors that identify stationary‑phase source subsets.
    An example with three cross-correlation windows and two source states is depicted, although their quantities can be chosen arbitrarily.
\label{fig:one}}
\end{figure}

We treat the realizations $\textstyle \left\{\mathbf{w}_{ij}^{(m)}\right\}_{m=1}^{M_{ij}}$ as independent samples of a random vector $\mathbf{w}_{ij}$; therefore, the empirical average 
\begin{equation}
      \widehat{\mathbb{E}}\left[ {\mathbf{w}}_{ij} \right]  =
\frac{1}{M_{ij}}
\sum_{m=1}^{M_{ij}} \mathbf{w}_{ij}^{(m)}
\label{eq:linear_average}
\end{equation}
of $M_{ij}$ time windows provides an estimator of expectation $\mathbb{E}[\mathbf{w}_{ij}]$.
Many authors have demonstrated that, in theory, the empirical average in Equation~\ref{eq:linear_average} converges to the interstation Green's function, provided the ambient wavefield is isotropic, uncorrelated, and equipartitioned~\citep{campillo2003long,wapenaar2004retrieving,wapenaar2006green,snieder2010imaging}.
Another important theoretical result is that, when the above conditions are satisfied, $\widehat{\mathbb{E}}\left[ {\mathbf{w}}_{ij} \right]$ should exhibit causal–acausal symmetry~\citep{wapenaar2004retrieving,snieder2004extracting}, representing equivalence between
waves traveling in both directions between the receivers.
In practice, however, this symmetry is rarely achieved~\citep{bensen2007processing, stehly2006study,yao2009analysis,tsai2009establishing} because real ambient noise sources are not isotropic (as shown in Figure~\ref{fig:one}a) with seasonal variations in their distribution and bursts of body wave energy~\citep{stehly2007traveltime,zhan2013spurious,delaney2017passive}. 

We write the complete cross-correlation vector $\mathbf{w}_{ij}$ as:
\begin{equation}
\mathbf{w}_{ij}
= \mathrm{concat}\!\left(
\mathrm{rev}(\prescript{-}{}{\mathbf{w}}_{ij}),
\prescript{0}{}{\mathbf{w}}_{ij},
\prescript{+}{}{\mathbf{w}}_{ij}
\right),
\label{eq:w_concat}
\end{equation}
where $\mathrm{concat}(\cdot)$ denotes the concatenation of two or more vectors, and $\mathrm{rev}(\cdot)$ denotes the reversal of the order of vector elements.
Here, $\prescript{+}{}{\mathbf{w}}_{ij}$ and $\prescript{-}{}{\mathbf{w}}_{ij}$ denote the vectors associated with the causal and acausal branches of the cross-correlation, respectively, each arranged in order of increasing lag magnitude. 
%
The lack of symmetry between the causal and acausal branches means that 
$\widehat{\mathbb{E}}\left[ \prescript{+}{}{\mathbf{w}}_{ij} \right] \neq \widehat{\mathbb{E}}\left[ \prescript{-}{}{\mathbf{w}}_{ij} \right]$. 
If this is the case, then interstation tomography using either of the branches is not justified, since the assumption of equipartitioning is violated. In such cases, the empirical averages are also sensitive to the source distribution and more complicated sensitivity kernels have to the used during tomography~\citep{fichtner2015source,ermert2017ambient}, which is computationally expensive. 


%

%


%



In this paper, we present a data-driven approach, termed Coherent Source Subsampling (CSS), which subsamples the set $\textstyle\left\{\mathbf{w}_{ij}^{(m)}\right\}$ based on the state of ambient noise sources. 
We denote the source state using a discrete variable $\lambda$, and two such source states are illustrated in Figure~\ref{fig:one}a, using dotted and dashed lines.
Our workflow for a given branch $b \in \{+,-\}$, receiver pair $(i, j)$ and source state $\lambda$ is very simple:
\begin{itemize}
      \item Identify windows associated with a particular source state $\lambda$, in which the cross-correlations due to several sources are coherent, to form the subset $\prescript{b}{\lambda}{\mathcal{C}}_{ij}$;     
      \item and averages only those windows to obtain
\begin{equation}
\widehat{\mathbb{E}}\left[ \prescript{b}{}{\mathbf{w}}_{ij} \mid \lambda \right]
= \frac{1}{\left|\prescript{b}{\lambda}{\mathcal{C}}_{ij}\right|}
\sum_{m \in \prescript{b}{\lambda}{\mathcal{C}}_{ij}}
\prescript{b}{}{\mathbf{w}}_{ij}^{(m)}.
\label{eq:conditional_average}
\end{equation}
\end{itemize}
Here, $\left|\cdot\right|$ denotes the cardinality of a set, and note that the 
subset $\prescript{b}{\lambda}{\mathcal{C}}_{ij}$ is not only pair specific but also branch specific, meaning the 
sources associated with a particular source state differ between causal and acausal branches (see Figure~\ref{fig:one}a).
A straightforward approach to conditional averaging is to partition the data according to observable external variables, such as season, month, or time of day. For example, separate averages for winter and summer months could be computed, expecting coherent cross-correlations due to ocean activity or storms within a season~\citep{yang2008characteristics,zhan2013spurious}. However, such conditioning strategies may not necessarily align with the actual coherence states in the ambient wavefield. 

We adopt the representation learning framework to identify coherent source states in a data-driven manner (as shown in Figure~\ref{fig:one}b). Specifically, we use a variational autoencoder to identify the subset of windows associated with a particular source state $\lambda$, rather than on predefined external variables. 
This idea is demonstrated using synthetic experiments for non-uniform source distributions with varying noise levels in the Supplementary Text~S1.
As an example of the source state, consider 
the stationary-phase zone for a given branch, where the sources coherently contribute to the inter-station response. In this paper, we label the source state for this zone as $\lambda = 1$. Another source state is illustrated in Figure~\ref{fig:one}b.
We show that CSS effectively restores causal–acausal symmetry when averaging cross-correlations confined to a particular source state.
Specifically, we achieve 
\begin{equation}
    \widehat{\mathbb{E}}\left[ \prescript{+}{}{\mathbf{w}}_{ij} \mid \lambda \right] \approx\widehat{\mathbb{E}}\left[ \prescript{-}{}{\mathbf{w}}_{ij} \mid \lambda \right]
\end{equation} for a choice of source state $\lambda$ and station pair $(i,j)$.
Although the causal and acausal branches are redundant in the sense that they contain the same information about the interstation Green's function, the achievement of causal–acausal symmetry is important for two reasons.
First, it validates the Green’s function approximation, which is a prerequisite for robust dispersion and travel-time measurements for tomography.
Second, symmetry ensures that there is no contamination from directional source effects and other transients.
To our knowledge, this is the first work that achieves the theoretically predicted high degree of 
causal–acausal symmetry in ambient noise cross-correlations even when the underlying source distribution is highly non-uniform and time-varying. 
We further show that the improved symmetry leads to more accurate surface-wave tomographic maps.
%
%
Finally, note that while CSS performs nonlinear averaging, similar to phase-weighted stacking by ~\cite{schimmel1997noise}, it is unique in the sense that it achieves causal–acausal symmetry.

\section{Data and Methods}
\subsection{Noise crosscorrelations}
We observed that the choice of the study region is less important to the CSS framework, which applies to any ambient noise dataset. Here, we chose to demonstrate the method by analyzing continuous ambient seismic data recorded in central California using $20$ stations from the Central California Seismic Experiment (CCSE) Transportable Array deployment~\citep{ccse2015}. This area poses a stringent challenge for the method because the ambient-noise wavefield is highly heterogeneous and strongly directional. The Pacific Ocean acts as a dominant and persistent source of microseismic energy, leading to a marked causal–acausal asymmetry in interstation correlations~\citep{retailleau2021towards}. In conventional ambient-noise tomography, the correlation branch associated with waves arriving from the Pacific is often preferentially used for dispersion measurements because the opposite branch is weakly illuminated~\citep{jiang2018rayleigh}. The selected stations form a west–east transect that extends from the Southern Coast Ranges (SCR), through the Rinconada (RF) and San Andreas (SAF) faults system, and into the Great Valley Basin (GVB) (Supplementary Figure~S$5$). Data were processed using standard ambient-noise interferometry procedures discussed in~\cite{bensen2007processing}. The detailed preprocessing steps are described in the Supplementary Text S$2$.

Before detailing the network architecture and subsampling procedure, we first provide a simple explanation of stationary zones, highlighting the limitation of the linear empirical average in Equation~\ref{eq:linear_average} and motivating the conditional average in Equation~\ref{eq:conditional_average}.

\subsection{Stationary-zone source state}
Ambient noise sources associated with a stationary zone are crucial, and both the causal and acausal branches each possess their own stationary zone.
In the presence of an isotropic source distribution, the contribution of stationary zones dominates $\mathbb{E}\left[ \mathbf{w}_{ij} \right]$, leading to causal-acausal symmetry.
%
In stationary zones, for a given source signature $\mathbf{s}$, the cross-correlation $\mathbf{w}_{ij}$ exhibits coherence across the noise sources.
We now give a simple ray-theoretical argument to show that these stationary zones correspond to regions where the travel-time difference between two stations is locally constant. 
The travel-time difference $\tau_{ij} = T_j(\mathbf{x}_s) - T_i(\mathbf{x}_s)$ represents the lag time at which a specific seismic phase appears in the cross-correlation function between stations $i$ and $j$. 
Here, for simplicity, we ignore the frequency dependence of the lag time, which corresponds to the differential travel time of a phase propagating from the source location $\mathbf{x}_s$ to the two receivers.
%




%
%
For branch $b$, the cross-correlation vector $\prescript{b}{}{\mathbf{w}}_{ij}$ depends on two random variables: the signature of the source $\mathbf{s}$ and the location of the source $\mathbf{x}_s$, whose joint distribution is denoted by $\rho(\mathbf{s},\mathbf{x}_s)$.
The distribution $\rho(\mathbf{s},\mathbf{x}_s)$ encapsulates the statistical properties of the ambient noise field, including source anisotropy and temporal variability.
The expectation of the random vector $\prescript{b}{}{\mathbf{w}}_{ij}$ is given by
      \begin{equation}
\mathbb{E}\big[\prescript{b}{}{\mathbf{w}}_{ij}\big]
= \iint \prescript{b}{}{\mathbf{w}}_{ij}\,
\rho(\mathbf{x}_s, \mathbf{s})\, \mathrm{d}\mathbf{x}_s\, \mathrm{d}\mathbf{s} = \iint \mathbb{E}\big[\prescript{b}{}{\mathbf{w}}_{ij} \mid \tau_{ij}\big]\,
\rho(\tau_{ij}, \mathbf{s})\, \mathrm{d}\tau_{ij}\,\mathrm{d}\mathbf{s},
\label{eq:causal_expectation}
\end{equation}
where $\mathbb{E}\big[\prescript{b}{}{\mathbf{w}}_{ij} \mid \tau_{ij}\big]$ is the conditional expectation of $\prescript{b}{}{\mathbf{w}}_{ij}$ given the difference in travel-time $\tau_{ij}$.
For a specific seismic phase arrival, we can transform the source distribution using the lag time $\tau_{ij} = T_j(\mathbf{x}_s) - T_i(\mathbf{x}_s)$.
The pushforward distribution $\rho(\tau_{ij}, \mathbf{s})$ is obtained from the joint distribution $\rho(\mathbf{x}_s, \mathbf{s})$ as:
\begin{equation}
\rho(\tau_{ij}, \mathbf{s}) =
\int \delta\big(\tau_{ij} - T_{ij}(\mathbf{x}_s)\big)\, 
\rho(\mathbf{x}_s, \mathbf{s})\, \mathrm{d}\mathbf{x}_s
= \int_{T_{ij}(\mathbf{x}_s) = \tau_{ij}}
\frac{\rho(\mathbf{x}_s, \mathbf{s})}{\|\nabla T_{ij}(\mathbf{x}_s)\|}\,
\mathrm{d}S(\mathbf{x}_s).
\label{eq:rho_pushforward}
\end{equation}
Here, integration is performed over a manifold,
where $\mathrm{d}S(\mathbf{x}_s)$ denotes the surface element on the manifold $T_{ij}(\mathbf{x}_s) = \mathrm{constant}$.
The gradient $\nabla T_{ij}(\mathbf{x}_s)$ measures how sensitively the
travel-time delay between receivers $i$ and $j$ changes with respect to the
source position $\mathbf{x}_s$. In stationary zones, small perturbations
in the source position do not affect the delay to the first order, which implies that the $\ell_2$ norm of the gradient $\|\nabla T_{ij}(\mathbf{x}_s)\|$ is close to zero.
%
Due to this, the cross-correlations are coherent across sources located in stationary zones, leading to constructive interference.
We will later leverage this coherence in the design of our neural network architecture.

Although the cross-correlations are coherent within stationary zones, in practice, the energy contribution from such zones does not necessarily dominate the ensemble average over all available windows (Equation~\ref{eq:linear_average}) --- this is because the source density $\rho(\mathbf{s}, \mathbf{x}_s)$ may be negligibly small in those regions.
In other words, as Equation~\ref{eq:rho_pushforward} indicates that the resulting distribution $\rho(\tau_{ij}, \mathbf{s})$ depends jointly on the mapping $T_{ij}(\mathbf{x}_s)$, and on the spatial source distribution $\rho(\mathbf{x}_s, \mathbf{s})$, the departure of $\rho(\mathbf{x}_s, \mathbf{s})$ from theoretical assumptions biases the interstation Green's function 
retrieval.
In such cases, the retrieval of Green’s functions can be improved by isolating and averaging only windows from stationary zones.
We fix $\lambda = 1$ to denote the source state of the stationary zone, which is the only state of interest in this study.
Additional source states may exist, namely alternative configurations of the ambient noise field, defined by sources with similar signatures that yield coherent cross-correlations between receiver pairs for a given arrival; we label these by $\lambda = 2,\ldots,K$. Figure~\ref{fig:one}a and the synthetic experiments (Supplementary Text~S$1$) provide representative examples of such states.

\subsection{Latent variable modeling}


To enable data-driven identification of coherent source subsets $\prescript{b}{\lambda}{\mathcal{C}}_{ij}$, we adopt a latent-variable modeling approach based on variational autoencoders (VAEs). 
VAEs provide a probabilistic framework for learning interpretable latent representations~\citep{kingma2013auto,doersch2016tutorial,bishop2023deep,prince2023understanding} from high-dimensional data. Within this framework, we introduce a discrete latent variable $\lambda$~\citep{morton2021scalable}
to represent source states and design the network architecture such that $\lambda$ correlates with physically meaningful source states, such as the stationary-zone state.
The design is based on deliberate architectural decisions: a permutation-invariant encoder that accumulates coherent information~\citep{bharadwaj2024extracting} across all windows for a given pair and a categorical posterior distribution over $\lambda$ that naturally groups coherent cross-correlation windows.
We emphasize that alternative network designs may exist; however, the focus of this paper is not to exhaustively explore all possible architectures, but rather to demonstrate that a carefully designed VAE can successfully solve the CSS problem by learning to distinguish stationary-zone windows from incoherent noise. The following subsections brief the generative and inference models that implement this approach.


\subsubsection{Generative model}
We begin with a generative model that mathematically describes how observed cross-correlations are synthesized from latent variables.
That is to say, we train a decoder network that generates each cross-correlation waveform $\prescript{b}{}{\mathbf{w}}_{ij}^{(m)}$
using three distinct latent-variable components --- they are highlighted in Figure~\ref{fig:one}b.
\begin{itemize}
\item \textbf{Coherent source features} $\left(\prescript{b}{\lambda}{\mathbf{g}}_{ij}\right)$. This component
represents the coherent contributions associated with a particular source state $\lambda$, branch $b$ and a station pair $(i,j)$. Coherency means that this component is invariant to the ordering of time windows associated with a specific pair and branch. This invariance is achieved by inferring $\prescript{b}{\lambda}{\mathbf{g}}_{ij}$ from the entire ensemble of $M$ cross-correlation windows for the station pair $(i,j)$ and the branch $b$.


\item \textbf{Probabilities of the source state} $\left(\prescript{b}{\lambda}{\pi}_{ij}^{(m)}\right)$.
This component represents the probability that the $m$th time window belongs to the source state $\lambda$ for the branch $b$ and the station pair $(i,j)$. These categorical weights satisfy
\begin{equation}
\sum_{\lambda=1}^{K} \prescript{b}{\lambda}{\pi}_{ij}^{(m)} = 1,
\end{equation}
and are used later to identify the windows associated with each source state.



\item \textbf{Nuisance features} $\left(\prescript{b}{}{\mathbf{n}}^{(m)}_{ij}\right)$.
This component is specific to each time window $m$ and captures features that are not linked with any of the source states --- for example, the cross-correlations due to incoherent sources away from the stationary zones.
\end{itemize}
In summary, our framework generates each cross-correlation waveform as:
\begin{equation}
\prescript{b}{}{\widehat{\mathbf{w}}}_{ij}^{(m)} = \mathtt{decoder}\left( \prescript{b}{l}{\mathbf{g}}_{ij}, \prescript{b}{}{\mathbf{n}}^{(m)}_{ij}\right),\quad \text{where}\; l = \arg\max_{\lambda} \prescript{b}{\lambda}{\pi}_{ij}^{(m)},\quad m=1,\ldots,M_{ij}.
\label{eq:generative_model}
\end{equation}
%
%
Here, $l$ denotes the source state with the highest posterior probability for the window $m$.
The $\mathtt{decoder}$ effectively selects a single coherent feature $\prescript{b}{l}{\mathbf{g}}_{ij}$ for generation of the observed cross-correlation.
%
%
We implement it as a convolutional neural network.



\subsubsection{Inference model}

The generative model in Equation~\ref{eq:generative_model} defines the forward process to synthesize cross-correlations 
from latent variables. To complete the VAE framework, 
we now detail the corresponding inference model that inverts this process by estimating the 
latent-variable components from observed data.
We perform probabilistic inference on the latent-variable components listed earlier using three components
 of the $\mathtt{encoder}$ network:
\begin{equation}
\begin{aligned}
\mathrm{concat}\left(\prescript{b}{1}{\mathbf{g}}_{ij},\ldots,\prescript{b}{K}{\mathbf{g}}_{ij}\right)
&= \mathtt{enc}_{\mathbf{g}}\left(\mathrm{concat}\left(\prescript{b}{}{\mathbf{w}}_{ij}^{(1)},\ldots,\prescript{b}{}{\mathbf{w}}_{ij}^{(M_{ij})}\right)\right),\\[2pt]
\mathrm{concat}\left(\prescript{b}{1}{\pi}_{ij}^{(m)},\ldots,\prescript{b}{K}{\pi}_{ij}^{(m)}\right)
&= \mathrm{softmax}\left(\mathtt{enc}_{\pi}(\prescript{b}{}{\mathbf{w}}_{ij}^{(m)})\right),\quad \text{for each}\; m=1,\ldots,M_{ij},\\[2pt]
\prescript{b}{}{\mathbf{n}}_{ij}^{(m)}
&= \mathtt{enc}_{\mathbf{n}}\left(\prescript{b}{}{\mathbf{w}}_{ij}^{(m)}\right),\quad \text{for each}\; m=1,\ldots,M_{ij}.
\end{aligned}
\end{equation}
Here, $\mathtt{enc}_{\mathbf{g}}$, $\mathtt{enc}_{\pi}$, and $\mathtt{enc}_{\mathbf{n}}$ are convolutional encoder components that infer coherent source features, source state probabilities, and nuisance features, respectively.
 The symmetric encoder component $\mathtt{enc}_{\mathbf{g}}$ is permutation invariant, i.e., its output $\prescript{b}{\lambda}{\mathbf{g}}_{ij}$ is invariant to the ordering of input windows.
  This component accumulates information across all time windows for a given station pair and branch.
 %
 On the contrary, $\mathtt{enc}_{\pi}$ and $\mathtt{enc}_{\mathbf{n}}$ operate on individual 
 windows to capture window-specific features, inferring $\prescript{b}{\lambda}{\pi}_{ij}^{(m)}$ and $\prescript{b}{}{\mathbf{n}}^{(m)}_{ij}$, respectively.
%
Note that the autoencoder may assign the stationary-zone state to any index in $\{1,\cdots,K\}$. After training, we therefore relabeled the source states so that the stationary-zone state is always set to $\lambda = 1$.
Additional architectural and training details are given in the Supplementary Text~S$3$.

\subsection{Coherent source subsampling (CSS)}

We employ an outer training loop over station pairs: for each pair $(i,j)$ we initialize the 
VAE from the previous section and train it jointly on both branches of that specific pair.
 After training, 
CSS aims to approximate the conditional expectation of each branch $b$ for pair $(i,j)$:
\begin{equation}
\mathbb{E}\left[ \prescript{b}{}{\mathbf{w}}_{ij} \mid \lambda \right]
= \iint
\prescript{b}{}{\mathbf{w}}_{ij}(\mathbf{x}_s,\mathbf{s})
\,\rho(\mathbf{x}_s,\mathbf{s} \mid \lambda)
\,\mathrm{d}\mathbf{x}_s\,\mathrm{d}\mathbf{s},
\label{eq:conditional_expectation}
\end{equation}
where the condition isolates contributions from sources associated with the state indexed by $\lambda$.
It uses the learned source state probabilities $\prescript{b}{\lambda}{\pi}_{ij}^{(m)}$ for subsampling.
Specifically, for each branch $b$, station pair $(i,j)$, and source state $\lambda$, the CSS estimator of the conditional expectation in Equation~\ref{eq:conditional_expectation} is given by Equation~\ref{eq:conditional_average}, 
where the subset of correlation windows $\prescript{b}{\lambda}{\mathcal{C}}_{ij}$ is defined as:     
\begin{equation}
      \prescript{b}{\lambda}{\mathcal{C}}_{ij} = \left\{ m : \prescript{b}{\lambda}{\pi}_{ij}^{(m)} > \alpha \right\}.
\label{eq:window_assignment}
\end{equation}
It is important to note that the identification of windows belonging to a specific source state is accomplished by probability thresholding (we used $\alpha = 0.85$).
By averaging only the time windows with high posterior probability $\prescript{b}{\lambda}{\pi}_{ij}^{(m)}$, we in effect subsample the windows associated with a given source state. 
With this targeted subsampling and by setting $\lambda = 1$, we can isolate the contributions from stationary regions in the source field.
Note that the conditional average $\widehat{\mathbb{E}}\left[ \prescript{b}{}{\mathbf{w}}_{ij} \mid \lambda \right]$ in Equation~\ref{eq:conditional_average} is a non‑linear function of the data, unlike linear averaging, because the subset $\prescript{b}{\lambda}{\mathcal{C}}_{ij}$ identification itself depends on the data.
It is important to emphasize that the averaging is performed on the original cross-correlation windows, using the VAE only to identify which windows belong to each source state.
Apart from the source state probabilities, we do not use any other outputs of the encoder or decoder networks, such as the reconstructed cross-correlations, in our analysis.
We now demonstrate the advantages of subsampling over indiscriminate averaging.

\section{Results}
\begin{figure}
\centering
\includegraphics[height=1.25\textwidth]{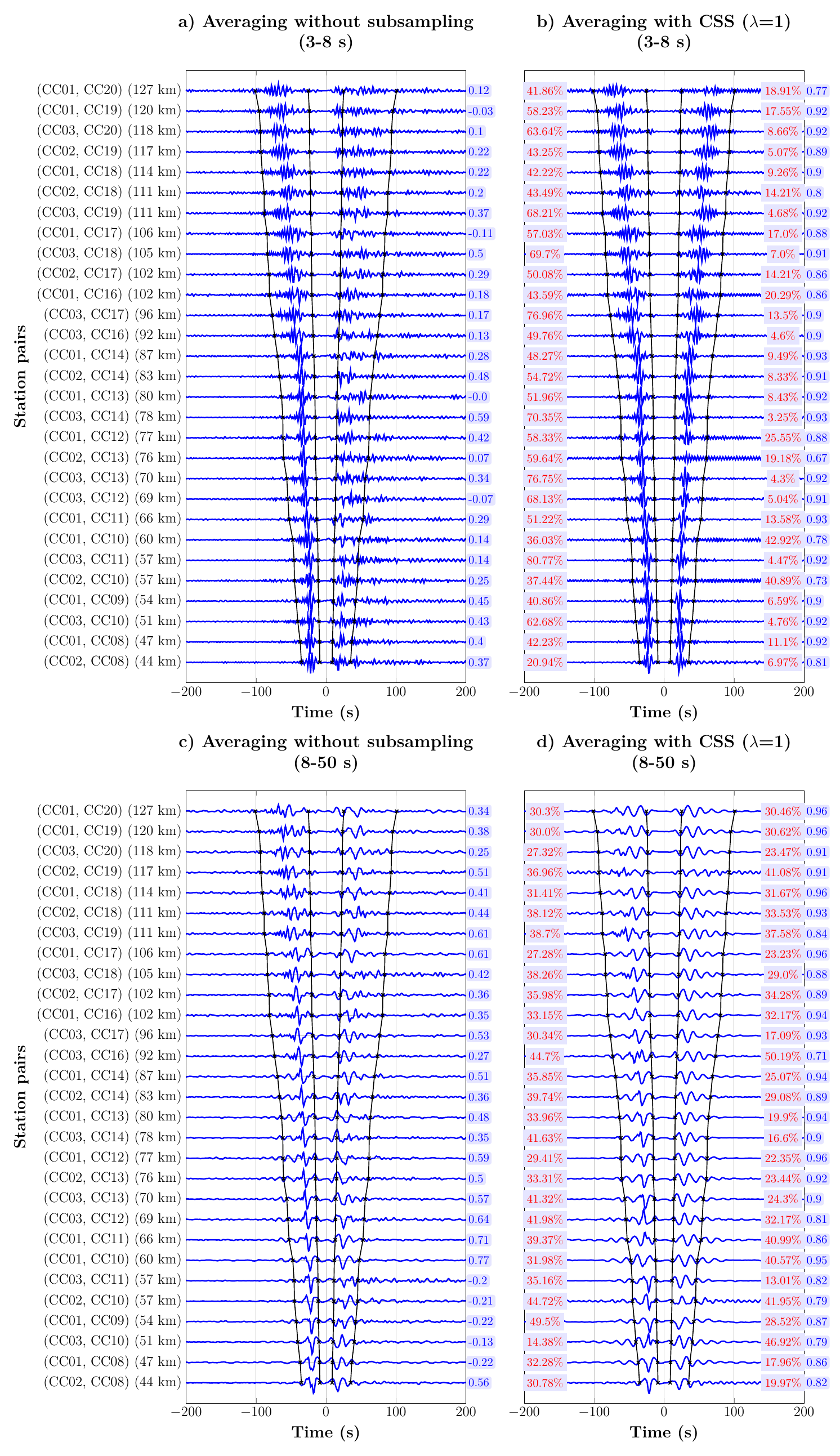}
\caption{Cross-correlation gathers from the Central California array. 
(a,c) Linear averaging without subsampling, $\widehat{\mathbb{E}}\left[ \prescript{b}{}{\mathbf{w}}_{ij} \right]$, for the $3$--$8\,$s and $8$--$50\,$s bands, respectively. The gathers exhibit clear causal--acausal asymmetry, with higher signal-to-noise ratios on the acausal side due to dominant illumination from Pacific Ocean microseisms. 
(b,d) The conditional averaging, $\widehat{\mathbb{E}}\left[ \prescript{b}{}{\mathbf{w}}_{ij} \mid \lambda=1 \right]$ using CSS for the same bands, shows improved symmetry 
between causal and acausal branches.
Red percentages denote the fraction of time windows assigned to the source state \text{$\lambda=1$} by CSS, and blue values show the correlation between causal and acausal branches for each station pair. The black lines mark time lags corresponding to apparent velocities of $1.25$ and $5\,$km/s.
}
\label{fig:gathers_Cal}
\end{figure}
%
For the Central California array, the correlations
after averaging without subsampling, $\widehat{\mathbb{E}}\left[ \prescript{b}{}{\mathbf{w}}_{ij} \right]$, shown in Figures~\ref{fig:gathers_Cal}a and \ref{fig:gathers_Cal}c, reveal a strongly directional ambient noise wavefield and display clear causal–acausal asymmetry. For both $3$--$8\,$s and $8$--$50\,$s period bands, the dominant energy is present in the acausal branch, consistent with seismic energy primarily originating from the Pacific Ocean. Because the acausal branch exhibits a signal-to-noise ratio (SNR) much higher than that of the causal branch, it is typically used for tomographic imaging. However, it should be noted that this measurement may be biased relative to the true interstation response, as indicated by the observed asymmetry.
In contrast, the conditionally averaged correlations (subsampled averages), $\widehat{\mathbb{E}}\left[ \prescript{b}{}{\mathbf{w}}_{ij} \mid \lambda=1 \right]$, shown in Figures~\ref{fig:gathers_Cal}b and\ref{fig:gathers_Cal}d exhibit a markedly stronger symmetry: the Rayleigh waves on the causal and acausal branches are nearly identical. This enhanced symmetry is quantified in Figure~\ref{fig:gathers_Cal}. 
For each pair of stations, 
Figure~\ref{fig:gathers_Cal} also reports the fraction $\frac{\left|\prescript{b}{\lambda}{\mathcal{C}}_{ij}\right|}{M_{ij}}$ of the retained time windows. The increased causal–acausal similarity indicates that CSS preferentially selects windows associated with the stationary source illumination state. We now obtained correlations that more accurately represent the true interstation response.

\begin{figure}
\centering
\noindent\includegraphics[width=1.1\columnwidth]{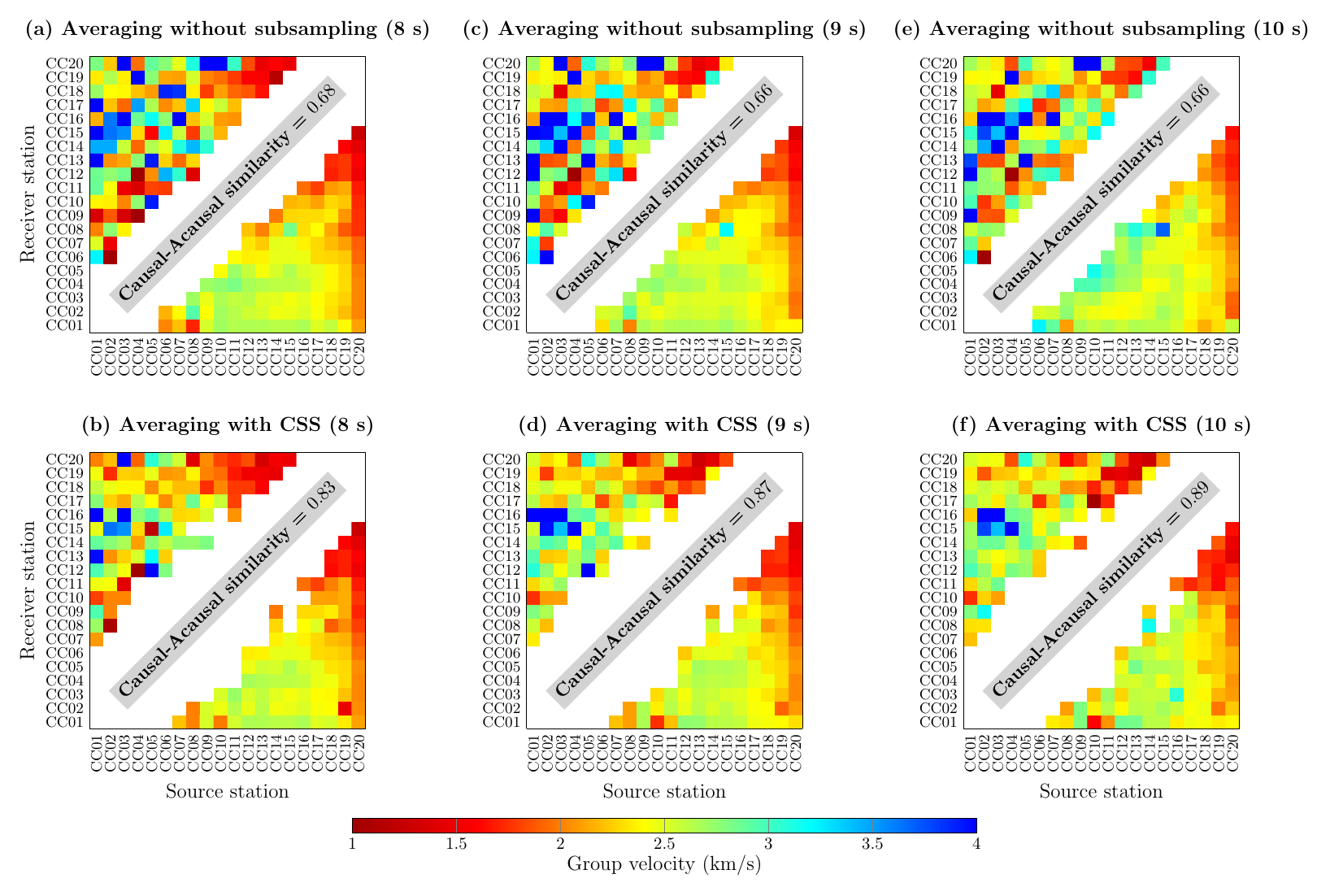}
\caption{Group‐velocity matrices obtained from ambient noise cross-correlations. For each subplot, the columns correspond to virtual source stations and rows to receiver stations (CC$01$--CC$20$). The color scale indicates Rayleigh-wave group velocity (km/s). Top row (a,c,e) shows picks without subsampling for periods of $8\,$s, $9\,$s, and $10\,$s, respectively, while the bottom row (b,d,f) shows the corresponding picks after Coherent Source Subsampling (CSS). White regions represent the lack of picks e.g., due to a smaller offset relative to Rayleigh-wave wavelength.
The gray diagonal annotation reports the average similarity between the causal and acausal branch picks.
Compared to linear averaging, CSS yields significantly higher reciprocity.
}
\label{fig:Dispersion_Matrix}
\end{figure}

In Figure~\ref{fig:Dispersion_Matrix}, we show that the
enhanced symmetry improves the dispersion measurements.
Here, the 
matrices derived without subsampling display pronounced artifacts: the group velocities vary irregularly between adjacent station pairs and across periods, particularly for the low-SNR branch (upper triangles of the matrices in Figures~\ref{fig:Dispersion_Matrix}a, c and e). 
Although the high-SNR group-velocity picks (lower triangles of Figures~\ref{fig:Dispersion_Matrix}a, c and e) do not exhibit significant artifacts, they may still be influenced by bias arising from non-uniform source illumination.
Our subsequent tomography findings will further support the presence of this bias.
In contrast to linear averaging, we observe that 
the CSS group-velocity picks are more similar across the branches --- note that the artifacts in the low-SNR branch are significantly reduced. 
In addition, the velocity picks in the high-SNR branch change slightly before and after subsampling, which reflects the reduction in bias.



\begin{figure}[ht]
\centering
\includegraphics[height=1.1\columnwidth,trim=10 0 0 0,clip]
{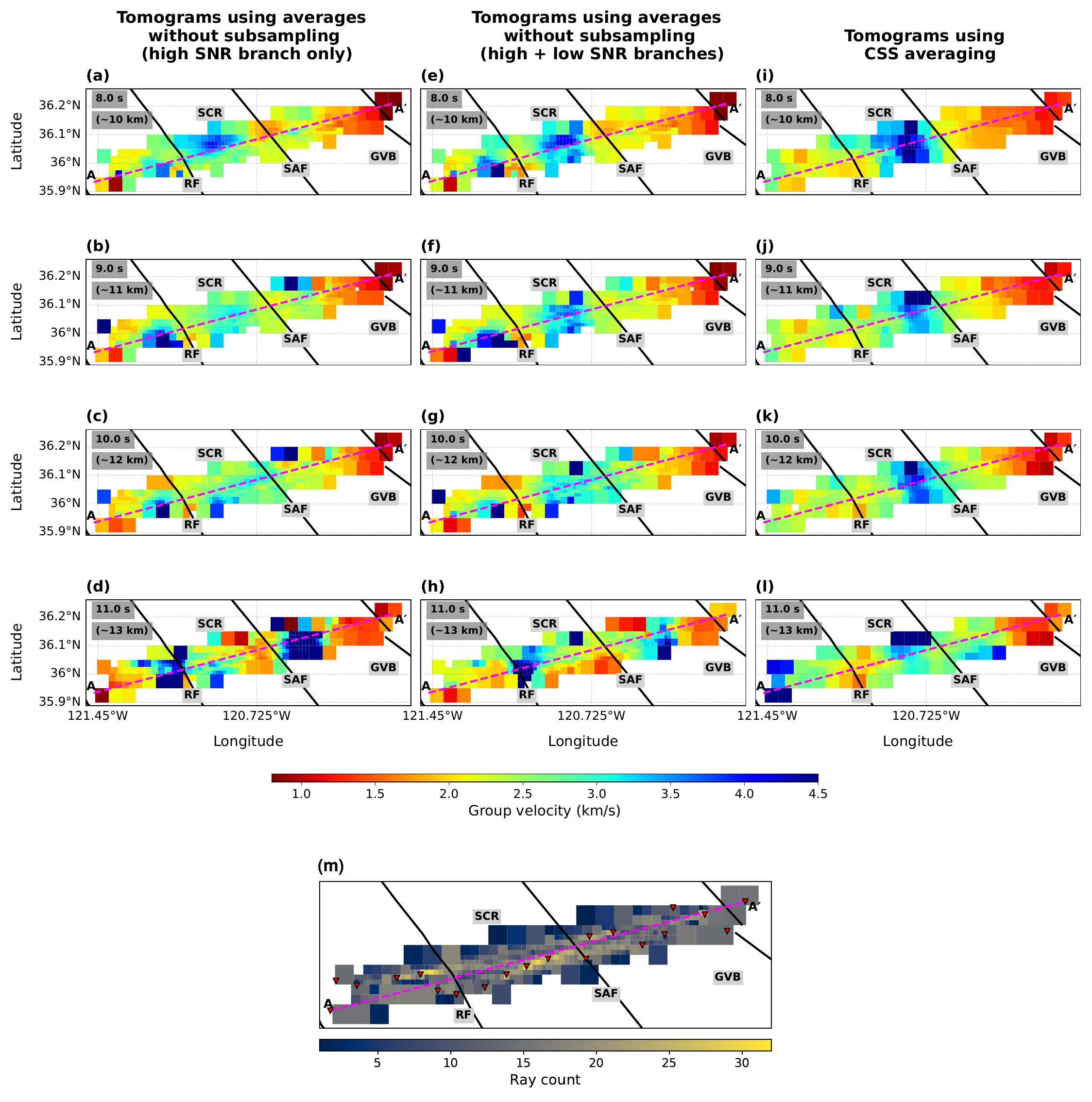}
\caption{Group‐velocity tomograms (a--l) and ray coverage (m) for the Central California array, where the stations (red inverted triangles) are roughly along the transect marked using the dashed magenta line. To obtain the 
tomograms for periods of $8$--$11\,$s (approximate peak sensitivity at depths of ~$10$--$13\,$km) the group-velocity picks are derived from: the high-SNR branch only (left column), both high- and low-SNR branches (middle column), and the proposed CSS selection (right column). Major tectonic elements, including the Southern Coast Ranges (SCR), Rinconada Fault (RF), San Andreas Fault (SAF), and Great Valley Basin (GVB), are indicated. 
The CSS tomograms mirror the underlying geology and reveal a high-velocity block located between the RF and SAF, a feature that is consistently supported by previous studies based on either refraction data or local-earthquake traveltimes.
%
}
\label{fig:Tomographs}
\end{figure}

Consequently, we construct three sets of group‐velocity  tomograms (see Figure~\ref{fig:Tomographs}) using:
 \begin{inparaenum}
    \item  picks from the high-SNR branch without subsampling --- hereafter, referred to as LS-high tomograms;
    \item  picks obtained without subsampling but after averaging the high- and low-SNR branches --- hereafter, referred to as LS tomograms;
    \item  picks derived from the correlations after CSS.
 \end{inparaenum}
For CSS tomography, since the group-velocity picks are generally consistent across the different branches, we averaged the values when they matched within a 15\% range and discarded the picks otherwise.
 %
 %
%
We used the surface‐wave tomography module of \textit{SeisLib}~\citep{magrini2022surface}, which performs a regularized least-squares tomography of interstation group‐velocity measurements.
 The regularization parameters were selected to yield comparable apparent smoothness and resolution in all cases so that the differences discussed below arise from the group-velocity picks themselves rather than from inversion settings. 
\begin{itemize}
    \item \textbf{Reduction of structural bias relative to fault geometry}: In the LS and LS-high models, velocity anomalies are elongated along the station array and approximately parallel to the dominant raypaths rather than to the mapped faults. Such along-array smearing reflects the bias due to non-uniform source distribution. In contrast, the CSS tomograms show velocity contrasts that follow the tectonic structure: distinct low-velocity zones adjacent to the RF and the SAF fault system, and a relatively higher-velocity block confined between them. This behavior is consistent with controlled-source and local-earthquake imaging studies near Parkfield, which document reduced velocities within the fault damage zone and comparatively higher velocities in adjacent crustal blocks~\citep{hole2006structure,yang20253} (Supplementary Figure S$7$). The improved alignment of anomalies with mapped structures indicates that CSS reduced source illumination bias and better recovers geologically consistent heterogeneity.
    \item \textbf{Improved consistency between periods and physical plausibility}: Surface-wave group velocities should evolve smoothly with period because sensitivity kernels broaden gradually with depth. The CSS models exhibit this expected behavior: the inter-fault block persists from ~$8\,$s to $~11\,$s (roughly $10$--$13\,$km depth) and weakens progressively with increasing period. By contrast, LS models display abrupt appearance and disappearance of anomalies and inconsistent velocity changes between neighboring periods. Such an inconsistency is unlikely to reflect real structure and instead suggests contamination from period-dependent source distribution, a known limitation of ambient-noise correlation under non-uniform illumination. Note that applying smoothing to reduce these inconsistencies will lead to a loss of resolution.
    
    \item \textbf{Shallow fault‐zone effects}: In shorter periods, CSS tomograms consistently resolve lower group velocities along the RF and SAF fault traces, while the intervening block remains faster. This pattern matches the expected seismic signature of a fault damage zone: fractured and fluid-altered rocks reduce seismic velocities within the fault core and surrounding damage region, while more intact crystalline rocks between faults maintain higher velocities. Similar low-velocity fault-zone structures have been reported from borehole and seismic studies at SAFOD and related Parkfield experiments~\citep{hole2006structure}. The LS tomograms either smear these regions along the array or fail to isolate them.
 
\end{itemize}
  These observations demonstrate that CSS improves ambient noise tomography by 
   recovering structurally coherent and geologically interpretable features.
We also processed another ambient-noise dataset~\citep{liang2008ambient} from the eastern United States, a stable cratonic region, to provide additional evidence; these results are presented in the Supplementary Text~S$4$.

\section{Discussion}


As CSS can effectively extract interstation response, it is particularly well suited for experiments using dense nodal arrays and distributed acoustic sensing (DAS). In these data-intensive settings, the goal of interferometry is to reduce the data volume while preserving useful subsurface information.
CSS can potentially be used to mitigate biases due to strong directional anthropogenic noise sources in these settings, improving interferometric extraction of interstation response.

The current practical limitation is that our implementation trains a separate VAE for each station pair; while this is manageable for arrays of moderate size, extending to hundreds of pairs would demand substantial computational resources. Development of new, joint multi-pair training schemes is essential as such scalable techniques are a key step toward applying CSS to dense seismic surveys.


\section{Conclusion}
We introduced Coherent Source Subsampling (CSS), a data-driven approach that reformulates ambient noise interferometry as a conditional averaging process guided by a neural-network based stationary-zone source selection. 
CSS suppresses the influence of transient and directionally biased noise and substantially enhances the causal–acausal symmetry of the retrieved cross-correlations.
Application to the central California dataset demonstrates that subsampling yields more reliable Rayleigh-wave arrivals, improved causal–acausal agreement, and more stable group-velocity dispersion curves across a range of interstation distances. 
These improvements reduce the bias and variability 
inherent to traditional ambient-noise tomography done without subsampling.
We demonstrated that the CSS tomograms successfully image established geological structures, including a high-velocity block located between the Rinconada and San Andreas faults, as well as low-velocity zones adjacent to these faults.
In summary, 
CSS extends the applicability of ambient noise interferometry to challenging environments with complex source distributions and shorter survey durations.

\section*{Open Research Section}

Waveform data used in this study can be obtained from the IRIS EarthScope Data Management Center 
(\href{https://ds.iris.edu/ds/nodes/dmc/}{https://ds.iris.edu/ds/nodes/dmc/}) 
through the FDSN dataselect service 
(\href{http://service.iris.edu/fdsnws/dataselect/1/}{http://service.iris.edu/fdsnws/dataselect/1/}). The dataset includes recordings from the CCSE array (network code TO)~\citep{ccse2015} and United States National Seismic Network (network code US)~\citep{USArray}. The codes developed for this study are included as a ZIP archive in the filetype ``Dataset" for peer review and will be made publicly available upon publication.

\section*{Conflict of Interest}
The authors declare that there are no conflicts of interest for this manuscript. 

\section*{Acknowledgments}
This work was supported by the Shell project under the project titled \emph{Neural Network Based Denoising Of Passive Seismic Data}.
We would like to thank Dr. Anu Chandran (Shell), 
Dr. Rene-Edouard Plessix (Shell), Dr. Arjun Datta (IISER Pune), Prof. Shyam Rai (IISER Pune), Prof. Aurélien Mordret (GEUS),
Prof. Gerard Schuster (University of Utah), and Dr. Kees Weemstra (KNMI) for their valuable feedback.
We also thank Prof. Brandon Schmandt (Rice University) for discussions on the CCSE dataset.
We thank Tiente Rengneichuong Koireng (IISc) for testing the method on other datasets. The authors used AI-assisted tools (OpenAI ChatGPT and Writefull) for language refinement. All edits were reviewed, and the authors are solely responsible for the final text.

%
%
\bibliographystyle{plainnat}
\bibliography{references}

@article{lin2008surface,
  title={Surface wave tomography of the western United States from ambient seismic noise: Rayleigh and Love wave phase velocity maps},
  author={Lin, Fan-Chi and Moschetti, Morgan P and Ritzwoller, Michael H},
  journal={Geophysical Journal International},
  volume={173},
  number={1},
  pages={281--298},
  year={2008},
  publisher={Blackwell Publishing Ltd Oxford, UK}
}

@article{yang2007ambient,
  title={Ambient noise Rayleigh wave tomography across Europe},
  author={Yang, Yingjie and Ritzwoller, Michael H and Levshin, Anatoli L and Shapiro, Nikolai M},
  journal={Geophysical Journal International},
  volume={168},
  number={1},
  pages={259--274},
  year={2007},
  publisher={Blackwell Publishing Ltd Oxford, UK}
}

@article{ritzwoller2011ambient,
  title={Ambient noise tomography with a large seismic array},
  author={Ritzwoller, Michael H and Lin, Fan-Chi and Shen, Weisen},
  journal={Comptes Rendus Geoscience},
  volume={343},
  number={8-9},
  pages={558--570},
  year={2011},
  publisher={Elsevier}
}

@article{tsai2009establishing,
  title={On establishing the accuracy of noise tomography travel-time measurements in a realistic medium},
  author={Tsai, Victor C},
  journal={Geophysical Journal International},
  volume={178},
  number={3},
  pages={1555--1564},
  year={2009},
  publisher={Blackwell Publishing Ltd Oxford, UK}
}

@article{stehly2006study,
  title={A study of the seismic noise from its long-range correlation properties},
  author={Stehly, L and Campillo, Michel and Shapiro, NM},
  journal={Journal of Geophysical Research: Solid Earth},
  volume={111},
  number={B10},
  year={2006},
  publisher={Wiley Online Library}
}

@article{fichtner2015source,
  title={Source-structure trade-offs in ambient noise correlations},
  author={Fichtner, Andreas},
  journal={Geophysical Journal International},
  volume={202},
  number={1},
  pages={678--694},
  year={2015},
  publisher={Oxford University Press}
}

@article{yang2008characteristics,
  title={Characteristics of ambient seismic noise as a source for surface wave tomography},
  author={Yang, Yingjie and Ritzwoller, Michael H},
  journal={Geochemistry, Geophysics, Geosystems},
  volume={9},
  number={2},
  year={2008},
  publisher={Wiley Online Library}
}

@article{zhan2013spurious,
  title={Spurious velocity changes caused by temporal variations in ambient noise frequency content},
  author={Zhan, Zhongwen and Tsai, Victor C and Clayton, Robert W},
  journal={Geophysical Journal International},
  volume={194},
  number={3},
  pages={1574--1581},
  year={2013},
  publisher={Oxford University Press}
}

@article{schimmel1997noise,
  title={Noise reduction and detection of weak, coherent signals through phase-weighted stacks},
  author={Schimmel, Martin and Paulssen, Hanneke},
  journal={Geophysical Journal International},
  volume={130},
  number={2},
  pages={497--505},
  year={1997},
  publisher={Blackwell Publishing Ltd Oxford, UK}
}

@misc{USArray,
  doi = {10.7914/SN/US},
  url = {https://www.fdsn.org/networks/detail/US/},
  author = {{Albuquerque Seismological Laboratory (ASL)/USGS}},
  title = {United States National Seismic Network},
  publisher = {International Federation of Digital Seismograph Networks},
  year = {1990}
}

@article{morton2021scalable,
  title={Scalable estimation of microbial co-occurrence networks with variational autoencoders},
  author={Morton, James T and Silverman, Justin and Tikhonov, Gleb and L{\"a}hdesm{\"a}ki, Harri and Bonneau, Rich},
  journal={BioRxiv},
  pages={2021--11},
  year={2021},
  publisher={Cold Spring Harbor Laboratory}
}

@article{ermert2017ambient,
  title={Ambient seismic source inversion in a heterogeneous earth: Theory and application to the Earth's hum},
  author={Ermert, Laura and Sager, Korbinian and Afanasiev, Michael and Boehm, Christian and Fichtner, Andreas},
  journal={Journal of Geophysical Research: Solid Earth},
  volume={122},
  number={11},
  pages={9184--9207},
  year={2017},
  publisher={Wiley Online Library}
}

@article{shapiro2004emergence,
  title={Emergence of broadband Rayleigh waves from correlations of the ambient seismic noise},
  author={Shapiro, Nikolai M and Campillo, Michel},
  journal={Geophysical Research Letters},
  volume={31},
  number={7},
  year={2004},
  publisher={Wiley Online Library}
}

@article{yao2009analysis,
  title={Analysis of ambient noise energy distribution and phase velocity bias in ambient noise tomography, with application to SE Tibet},
  author={Yao, Huajian and Van Der Hilst, Robert D},
  journal={Geophysical Journal International},
  volume={179},
  number={2},
  pages={1113--1132},
  year={2009},
  publisher={Blackwell Publishing Ltd Oxford, UK}
}

@article{liang2008ambient,
  title={Ambient seismic noise tomography and structure of eastern {North America}},
  author={Liang, Chuntao and Langston, Charles A},
  journal={Journal of Geophysical Research: Solid Earth},
  volume={113},
  number={B3},
  year={2008},
  publisher={Wiley Online Library}
}

@article{wang2025high,
  title={High-performance CPU-GPU heterogeneous computing method for 9-component ambient noise cross-correlation},
  author={Wang, Jingxi and Wang, Weitao and Wu, Chao and Jiang, Lei and Zou, Hanwen and Yao, Huajian and Chen, Ling},
  journal={Earthquake Research Advances},
  pages={100357},
  year={2025},
  publisher={Elsevier}
}

@article{bharadwaj2022redatuming,
  title={Redatuming physical systems using symmetric autoencoders},
  author={Bharadwaj, Pawan and Li, Matthew and Demanet, Laurent},
  journal={Physical Review Research},
  volume={4},
  number={2},
  pages={023118},
  year={2022},
  publisher={APS}
}

@article{bharadwaj2024extracting,
  title={On extracting coherent seismic wavefield using variational symmetric autoencoders},
  author={Bharadwaj, Pawan},
  journal={arXiv preprint arXiv:2411.15613},
  year={2024}
}

@article{doersch2016tutorial,
  title={Tutorial on variational autoencoders},
  author={Doersch, Carl},
  journal={arXiv preprint arXiv:1606.05908},
  year={2016}
}

@book{prince2023understanding,
  title={Understanding deep learning},
  author={Prince, Simon JD},
  year={2023},
  publisher={MIT press}
}

@article{bajad2025symmetric,
  title={Symmetric autoencoders for retrieving void-generated ringing modes and backscatter from vehicle seismic noise},
  author={Bajad, Sanket Narayan and Bharadwaj, Pawan},
  journal={Geophysics},
  volume={90},
  number={5},
  pages={1--69},
  year={2025},
  publisher={Society of Exploration Geophysicists}
}

@article{kingma2013auto,
  title={Auto-encoding variational bayes},
  author={Kingma, Diederik P and Welling, Max},
  journal={arXiv preprint arXiv:1312.6114},
  year={2013}
}

@book{bishop2023deep,
  title={Deep learning: Foundations and concepts},
  author={Bishop, Christopher M and Bishop, Hugh},
  year={2023},
  publisher={Springer Nature}
}

@article{snieder2004extracting,
  title={Extracting the Green’s function from the correlation of coda waves: A derivation based on stationary phase},
  author={Snieder, Roel},
  journal={Physical review E},
  volume={69},
  number={4},
  pages={046610},
  year={2004},
  publisher={APS}
}

@article{campillo2003long,
  title={Long-range correlations in the diffuse seismic coda},
  author={Campillo, Michel and Paul, Anne},
  journal={Science},
  volume={299},
  number={5606},
  pages={547--549},
  year={2003},
  publisher={American Association for the Advancement of Science}
}

@article{wapenaar2004retrieving,
  title={Retrieving the elastodynamic Green's function of an arbitrary inhomogeneous medium by cross correlation},
  author={Wapenaar, Kees},
  journal={Physical review letters},
  volume={93},
  number={25},
  pages={254301},
  year={2004},
  publisher={APS}
}

@preamble{"\newcommand{\SortNoop}[1]{}"}

@article{bensen2007processing,
  title={Processing seismic ambient noise data to obtain reliable broad-band surface wave dispersion measurements},
  author={Bensen, GD and Ritzwoller, MH and Barmin, MP and Levshin, A Lin and Lin, Feifan and Moschetti, MP and Shapiro, NM and Yang, Yanyan},
  journal={Geophysical journal international},
  volume={169},
  number={3},
  pages={1239--1260},
  year={2007},
  publisher={Blackwell Publishing Ltd Oxford, UK}
}

@article{trampert2003global,
  title={Global anisotropic phase velocity maps for fundamental mode surface waves between 40 and 150 s},
  author={Trampert, Jeannot and Woodhouse, John H},
  journal={Geophysical Journal International},
  volume={154},
  number={1},
  pages={154--165},
  year={2003},
  publisher={Blackwell Publishing Ltd Oxford, UK}
}

@article{ritzwoller1998eurasian,
  title={Eurasian surface wave tomography: Group velocities},
  author={Ritzwoller, Michael H and Levshin, Anatoli L},
  journal={Journal of Geophysical Research: Solid Earth},
  volume={103},
  number={B3},
  pages={4839--4878},
  year={1998},
  publisher={Wiley Online Library}
}

@article{ritzwoller2001crustal,
  title={Crustal and upper mantle structure beneath Antarctica and surrounding oceans},
  author={Ritzwoller, Michael H and Shapiro, Nikolai M and Levshin, Anatoli L and Leahy, Garrett M},
  journal={Journal of Geophysical Research: Solid Earth},
  volume={106},
  number={B12},
  pages={30645--30670},
  year={2001},
  publisher={Wiley Online Library}
}

@article{shapiro2002monte,
  title={Monte-Carlo inversion for a global shear-velocity model of the crust and upper mantle},
  author={Shapiro, NM and Ritzwoller, MH},
  journal={Geophysical Journal International},
  volume={151},
  number={1},
  pages={88--105},
  year={2002},
  publisher={Blackwell Publishing Ltd Oxford, UK}
}

@article{ekstrom2011global,
  title={A global model of Love and Rayleigh surface wave dispersion and anisotropy, 25-250 s},
  author={Ekstr{\"o}m, G{\"o}ran},
  journal={Geophysical Journal International},
  volume={187},
  number={3},
  pages={1668--1686},
  year={2011},
  publisher={Blackwell Publishing Ltd Oxford, UK}
}

@article{mordret2013near,
  title={Near-surface study at the Valhall oil field from ambient noise surface wave tomography},
  author={Mordret, A and Land{\`e}s, M and Shapiro, NM and Singh, SC and Roux, P and Barkved, OI},
  journal={Geophysical Journal International},
  volume={193},
  number={3},
  pages={1627--1643},
  year={2013},
  publisher={Oxford University Press}
}

@article{snieder2010imaging,
  title={Imaging with ambient noise},
  author={Snieder, Roel and Wapenaar, Kees},
  journal={Physics Today},
  volume={63},
  number={9},
  pages={44--49},
  year={2010},
  publisher={AIP Publishing}
}

@article{wapenaar2006green,
  title={Green’s function representations for seismic interferometry},
  author={Wapenaar, Kees and Fokkema, Jacob},
  journal={Geophysics},
  volume={71},
  number={4},
  pages={SI33--SI46},
  year={2006},
  publisher={Society of Exploration Geophysicists}
}

@article{stehly2007traveltime,
  title={Traveltime measurements from noise correlation: stability and detection of instrumental time-shifts},
  author={Stehly, L and Campillo, M and Shapiro, NM},
  journal={Geophysical Journal International},
  volume={171},
  number={1},
  pages={223--230},
  year={2007},
  publisher={Blackwell Publishing Ltd Oxford, UK}
}

@article{sabra2005surface,
  title={Surface wave tomography from microseisms in Southern California},
  author={Sabra, Karim G and Gerstoft, Peter and Roux, Philippe and Kuperman, WA and Fehler, Michael C},
  journal={Geophysical Research Letters},
  volume={32},
  number={14},
  year={2005},
  publisher={Wiley Online Library}
}

@article{yang2008structure,
  title={Structure of the crust and uppermost mantle beneath the western United States revealed by ambient noise and earthquake tomography},
  author={Yang, Yingjie and Ritzwoller, Michael H and Lin, F-C and Moschetti, MP and Shapiro, Nikolai M},
  journal={Journal of Geophysical Research: Solid Earth},
  volume={113},
  number={B12},
  year={2008},
  publisher={Wiley Online Library}
}

@article{delaney2017passive,
  title={Passive seismic monitoring with nonstationary noise sources},
  author={Delaney, Evan and Ermert, Laura and Sager, Korbinian and Kritski, Alexander and Bussat, Sascha and Fichtner, Andreas},
  journal={Geophysics},
  volume={82},
  number={4},
  pages={KS57--KS70},
  year={2017},
  publisher={Society of Exploration Geophysicists}
}

@misc{ccse2015,
  author       = {CCSE},
  year         = 2015,
  title        = {{Central California seismic experiment, caltech, Other/Seismic Network}},
  publisher    = {Caltech},
  doi          = {10.7909/C3B56GVW}
}

@article{jiang2018rayleigh,
  title={Rayleigh and S wave tomography constraints on subduction termination and lithospheric foundering in central California},
  author={Jiang, Chengxin and Schmandt, Brandon and Hansen, Steven M and Dougherty, Sara L and Clayton, Robert W and Farrell, Jamie and Lin, Fan-Chi},
  journal={Earth and Planetary Science Letters},
  volume={488},
  pages={14--26},
  year={2018},
  publisher={Elsevier}
}

@article{retailleau2021towards,
  title={Towards structural imaging using seismic ambient field correlation artefacts},
  author={Retailleau, Lise and Beroza, Gregory C},
  journal={Geophysical Journal International},
  volume={225},
  number={2},
  pages={1453--1465},
  year={2021},
  publisher={Oxford University Press}
}

@article{magrini2022surface,
  title={Surface-wave tomography using SeisLib: a Python package for multiscale seismic imaging},
  author={Magrini, Fabrizio and Lauro, Sebastian and K{\"a}stle, Emanuel and Boschi, Lapo},
  journal={Geophysical Journal International},
  volume={231},
  number={2},
  pages={1011--1030},
  year={2022},
  publisher={Oxford University Press}
}

@article{yang20253,
  title={3-D variational inference-based double-difference seismic tomography method and application to the SAFOD site, California},
  author={Yang, Hao and Zhang, Xin and Zhang, Haijiang},
  journal={Geophysical Journal International},
  volume={241},
  number={1},
  pages={378--404},
  year={2025},
  publisher={Oxford University Press}
}

@article{hole2006structure,
  title={Structure of the San Andreas fault zone at SAFOD from a seismic refraction survey},
  author={Hole, JA and Ryberg, Trond and Fuis, GS and Bleibinhaus, F and Sharma, AK},
  journal={Geophysical Research Letters},
  volume={33},
  number={7},
  year={2006},
  publisher={Wiley Online Library}
}

\end{document}


\maketitle


\noindent\textbf{Contents of this file}
\begin{enumerate}
\item Text S1 to S4
\item Figures S1 to S12
\end{enumerate}

\section*{Introduction}

\begin{figure}
\centering
\noindent\includegraphics[width=0.975\columnwidth]{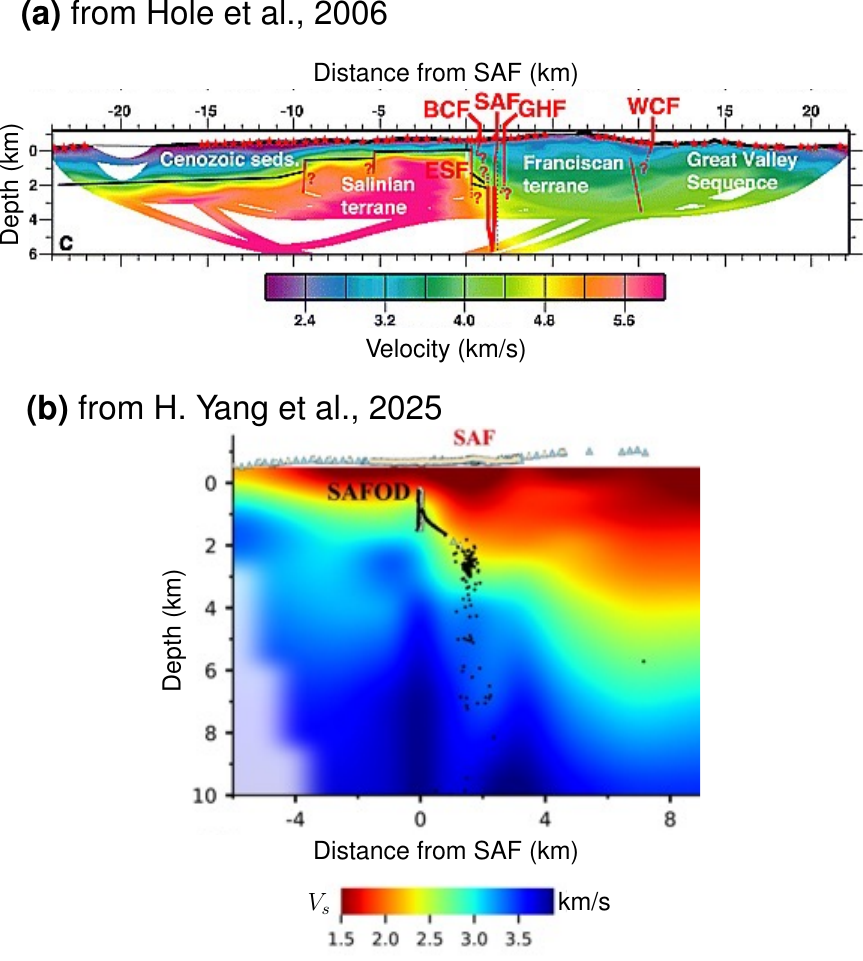}
\caption{Previously published seismic velocity cross-sections across the San Andreas Fault (SAF) near Parkfield used as an independent geological reference. (a) Shear-wave velocity model from earthquake and controlled-source imaging showing a pronounced low-velocity fault-zone damage region and higher velocities in adjacent crustal blocks (after~\cite{hole2006structure}). (b) Near-fault velocity structure derived from recent high-resolution seismic imaging around the SAFOD site, again revealing reduced velocities localized near the fault core (after~\cite{yang20253}).}
\label{fig:ref_tomo}
\end{figure}

This file provides supplementary methodological details and validation experiments supporting the results presented in the main manuscript. Text~S$1$ describes the synthetic experiments used to evaluate CSS, and Figures~\ref{fig:synthetic_data-1}--\ref{fig:synthetic_data-4} present the corresponding results. Figure~\ref{fig:map_California} shows the station distribution for the California data set, and Text~S$2$ details the characteristics of the dataset and the preprocessing workflow. Text~S$3$ outlines the SymVAE framework and its implementation, with the full architecture shown in Figure~\ref{fig:SymVAE}. Figure~\ref{fig:ref_tomo} presents previously published seismic velocity cross-sections across the San Andreas Fault used as an independent geological reference for comparison with the tomographic results. Finally, Text~S$4$ describes the application of the framework to the eastern United States data.

\section*{Text S1. Synthetic data}

\begin{figure}
\centering
\noindent\includegraphics[width=0.8\columnwidth]{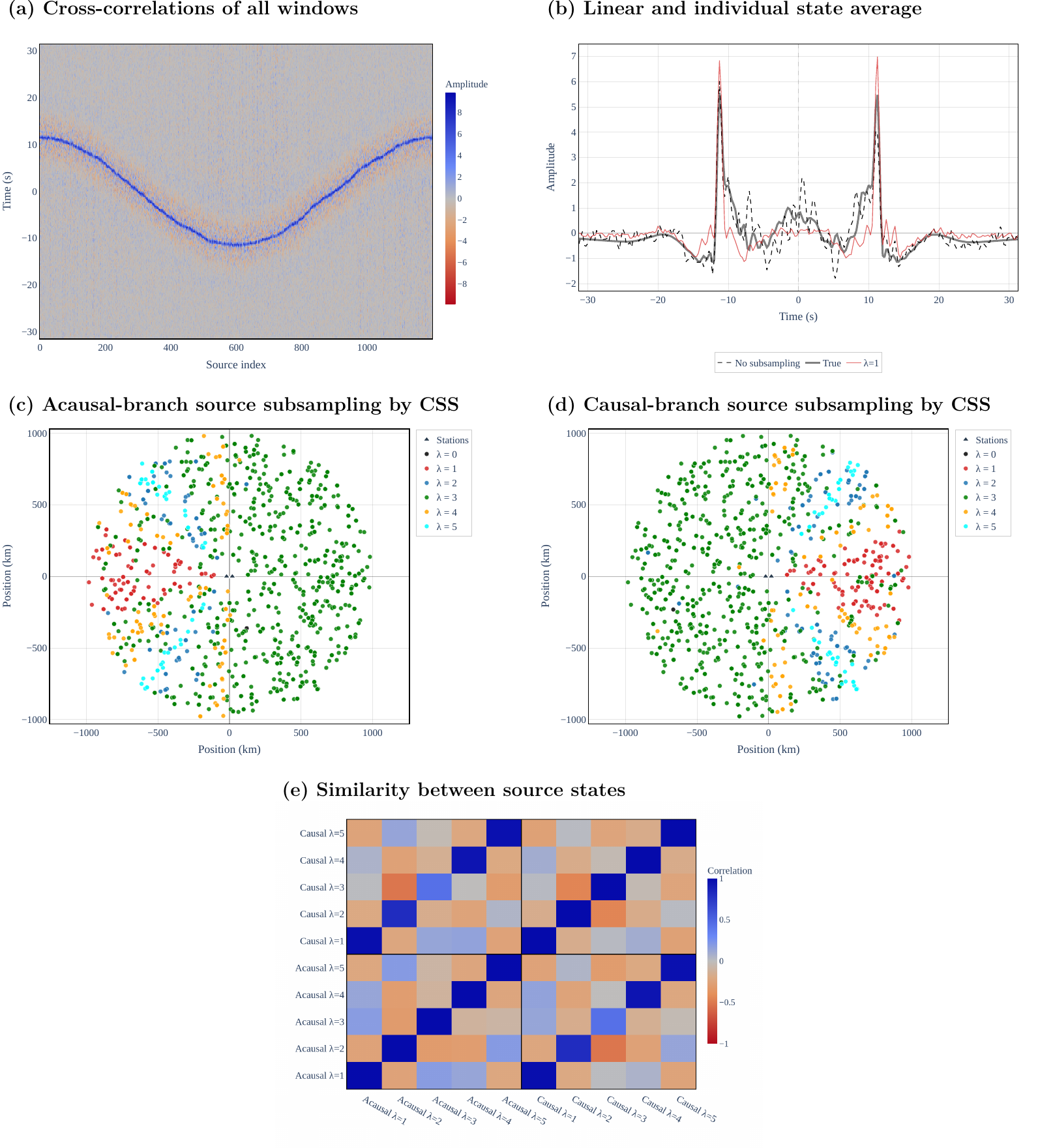}
\caption{Synthetic-data subsampling using CSS. (a) Inter-station cross-correlations for all synthetic noise windows. (b) Linear average of randomly sampled sources (No subsampling, black dashed), all sources denoted as true (gray), and source state for $\lambda\text{=1}$ (red). (c) and (d) Acausal and causal branches source locations corresponding to each source state. (e) Similarity matrix showing inter-state correlations between acausal and causal branches. The strongest similarity occurs for $\lambda\text{=1}$, indicating that CSS effectively isolates the most coherent source-state contribution.}
\label{fig:synthetic_data-1}
\end{figure}

\begin{figure}
\centering
\noindent\includegraphics[width=0.8\columnwidth]{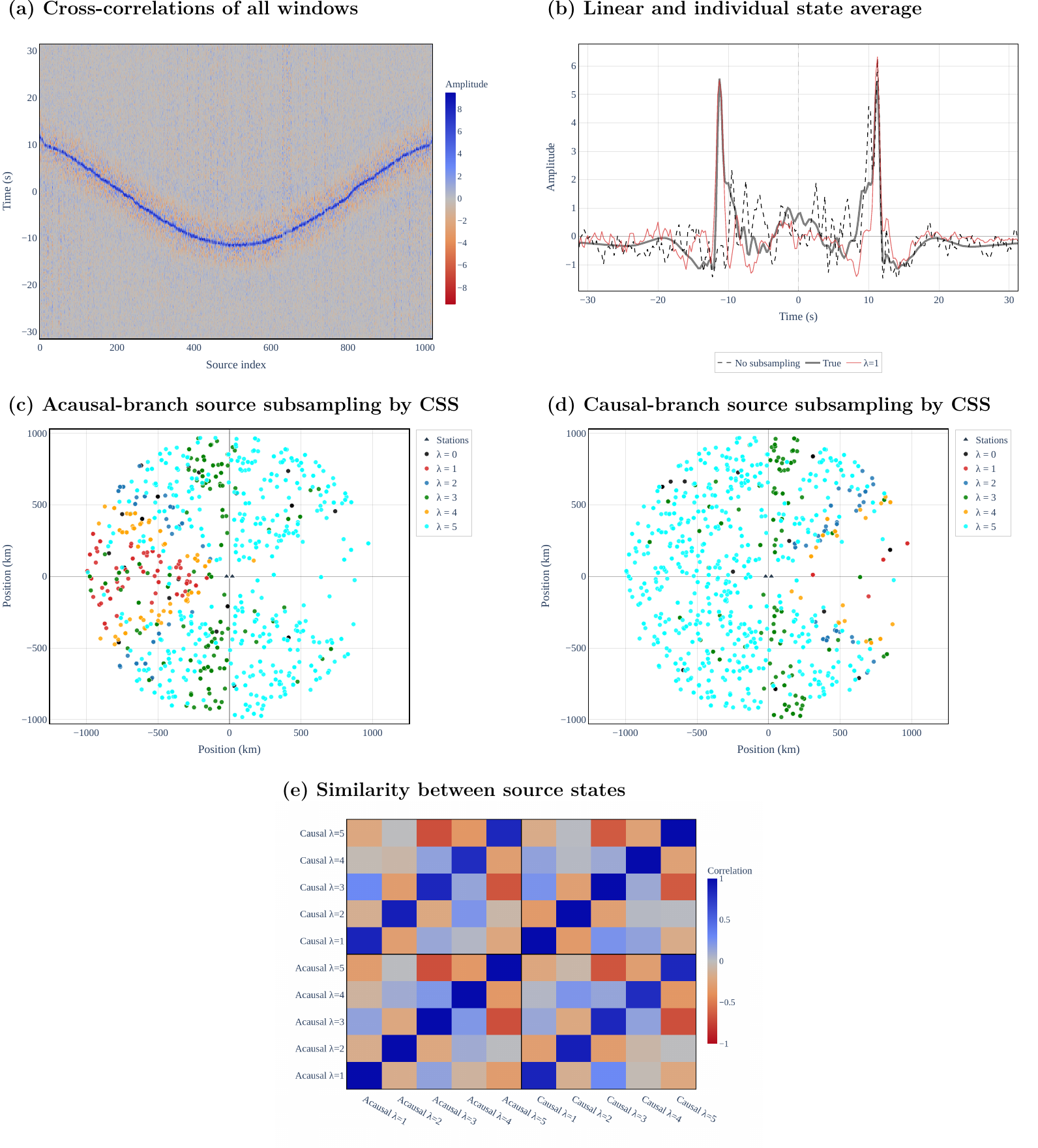}
\caption{CSS-based subsampling for non-uniformly distributed sources with sparse causal stationary-zone coverage. The causal branch contains fewer sources within the stationary zone. The strongest similarity occurs for $\lambda\text{=1}$, indicating that CSS effectively isolates the most coherent source-state contribution despite non-uniform source distribution.}
\label{fig:synthetic_data-2}
\end{figure}

\begin{figure}
\centering
\noindent\includegraphics[width=0.8\columnwidth]{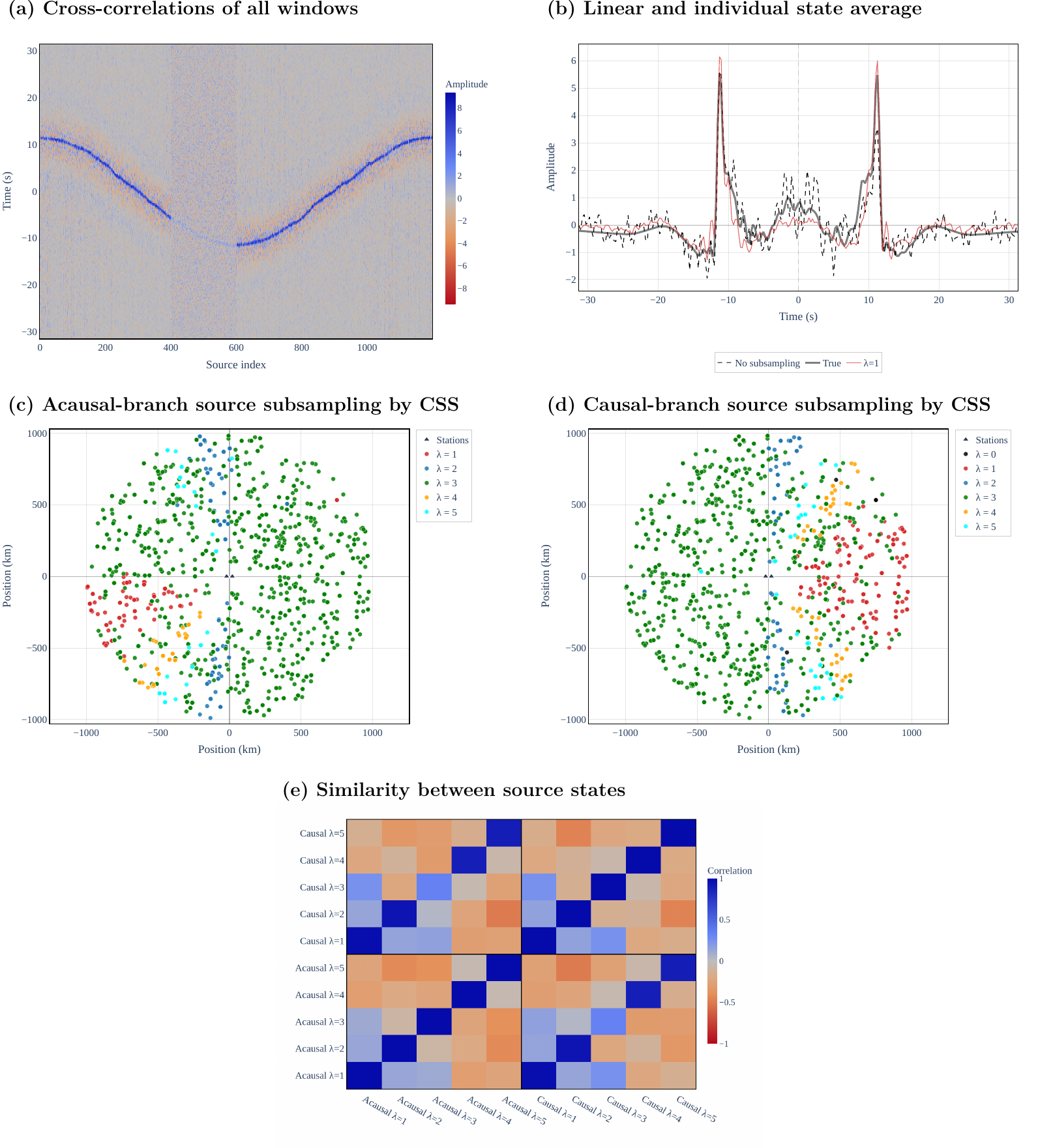}
\caption{Robustness of CSS towards higher noise level zones in uniformly distributed sources. A subset of acausal sources has higher noise levels. The strongest similarity occurs for $\lambda\text{=1}$. The results demonstrate that CSS is robust to noise and effectively ignores windows lacking coherent signal.}
\label{fig:synthetic_data-3}
\end{figure}

\begin{figure}
\centering
\noindent\includegraphics[width=0.8\columnwidth]{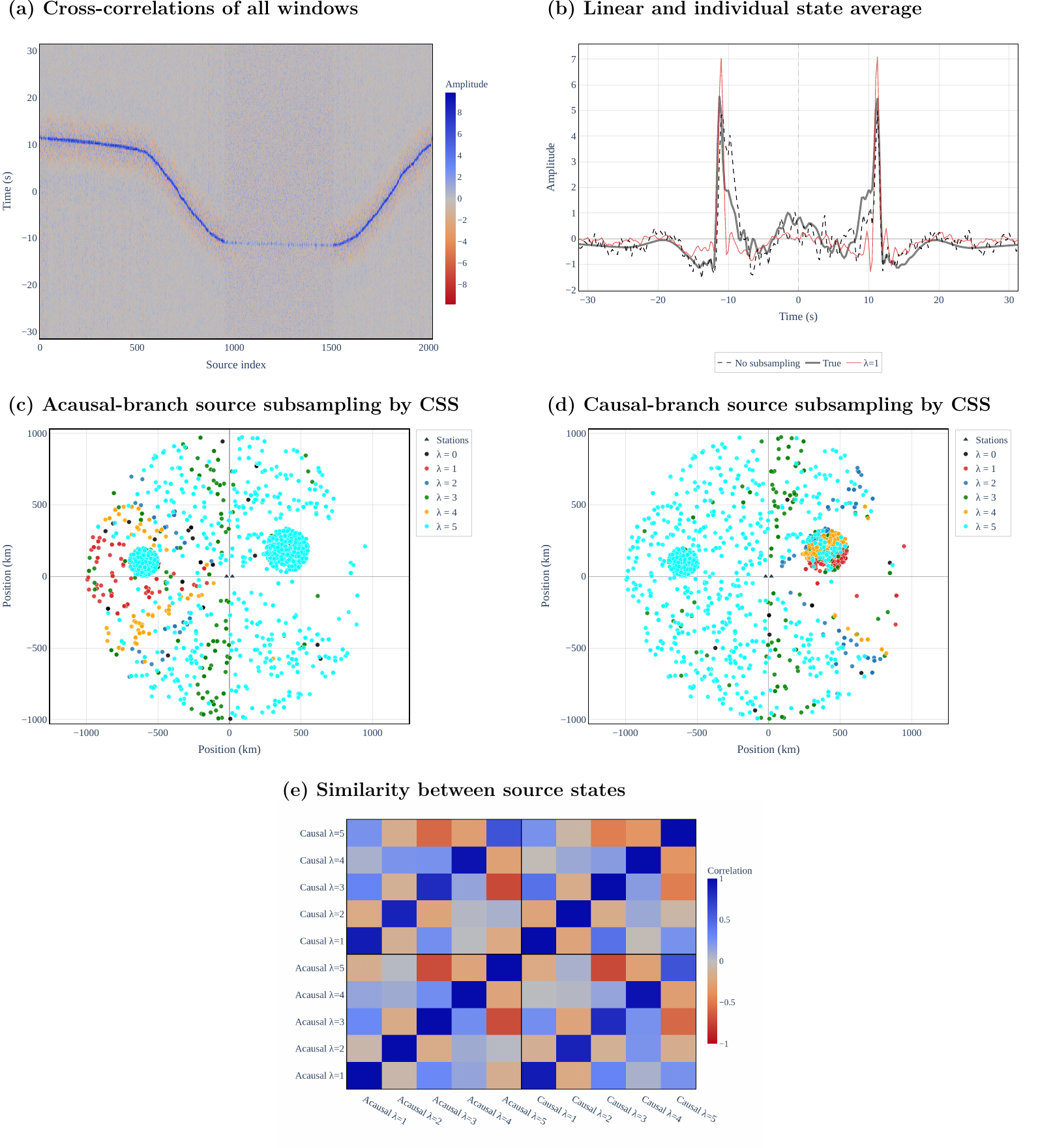}
\caption{CSS performance for complex non-uniform source distributions with mixed noise levels. Acausal and causal branch source distributions includes two dense source patches: one characterized by high noise in the acausal branch and another with low noise levels in causal branch. The strongest similarity occurs for $\lambda\text{=1}$, indicating that CSS effectively isolates the most coherent source-state contribution.}
\label{fig:synthetic_data-4}
\end{figure}

To analyze CSS behavior under controlled conditions, we construct synthetic ambient-noise data with sources distributed in all directions. We then averaged over all synthetic sources to obtain the \emph{True} cross-correlations, which serve as our reference. To mimic realistic noisy conditions, we select cross-correlation windows corresponding to some non-uniform source distribution from the full synthetic dataset; the linear stack of these windows corresponds to the \emph{No subsampling} trace in Figures~\ref{fig:synthetic_data-1}--\ref{fig:synthetic_data-4}. Figures~\ref{fig:synthetic_data-1}--\ref{fig:synthetic_data-4} show four different cases (i.e., source distributions with varying noise levels) for which this CSS analysis is performed. CSS is subsequently applied to this reduced dataset to recover the underlying source-state clusters. For each recovered source state $\lambda$, we compute the average cross-correlation for the acausal and causal branches separately. 
In particular, in all cases (Figures ~\ref{fig:synthetic_data-1}--\ref{fig:synthetic_data-4}), the subsampled average for the source state $\lambda\text{=1}$ shows the greatest acausal–causal consistency, and its waveform matches more closely the \emph{True}  correlation than the simple linear average of the selected windows. This indicates that CSS can isolate the most physically coherent source-state contribution, even when the data are sparse and non-uniformly distributed. To explain why $\lambda\text{=1}$ produces the most stable waveform in each case, we inspect the spatial distribution of sources assigned to each recovered state (Figures~\ref{fig:synthetic_data-1}c,d -- \ref{fig:synthetic_data-4}c,d). For both the acausal and causal branches, the sources associated with $\lambda\text{=1}$ are concentrated primarily within the stationary zones of the pair. In contrast, higher-index source states ($\lambda\text{=\{2,3,4,5\}}$) exhibit more dispersed and asymmetric source patterns, which explains their reduced contribution to a coherent acausal–causal reconstruction. The similarity matrices (Figure~\ref{fig:synthetic_data-1}e--\ref{fig:synthetic_data-4}e) further corroborate this interpretation: the acausal and causal signatures of the source state $\lambda\text{=1}$ have the highest mutual correlation, while the remaining states show a much weaker cross-branch similarity. Collectively, these findings across all four cases demonstrate that CSS effectively isolates the physically meaningful source-state in synthetic data, thereby improving correlation symmetry even when only limited, non-uniformly distributed sources are available.

\section*{Text S2. Central California dataset and preprocessing}

\begin{figure}[ht!]
\centering
\noindent\includegraphics[width=1.01\columnwidth]{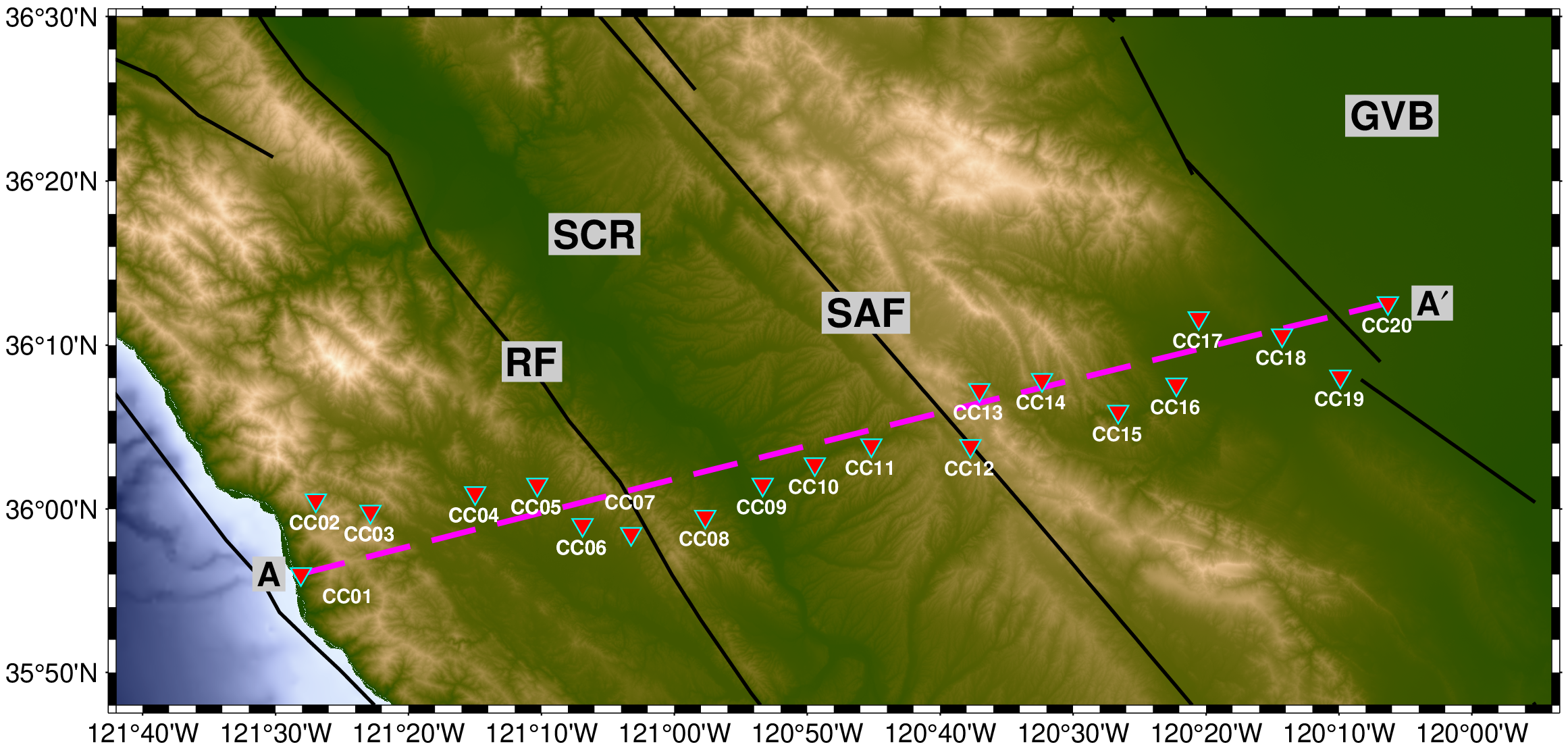}
\caption{Map of topography and station distribution in the study area. Stations of the TO network (CC$01$--CC$20$) are shown as red inverted triangles. The magenta line marks the location of cross-section A-A$'$ along which the directional cross-correlation results are presented. Black lines denote major mapped fault traces. Geologic abbreviations: SCR, Southern Coast Ranges; GVB, Great Valley Basin; SAF, San Andreas Fault; RF, Rinconada Fault.}
\label{fig:map_California}
\end{figure}

Figure~\ref{fig:map_California} shows the station distribution for the Central California dataset. The array consists of $20$ stations from the Central California Seismic Experiment (CCSE) Transportable Array that form a west–east transect that crosses the Southern Coast Ranges, the Rinconada Fault (RF), the San Andreas Fault (SAF), and the Great Valley Basin (GVB). The geometry intentionally samples strong lateral structural contrasts and regions of known directional ambient-noise illumination.

Continuous vertical-component recordings (HHZ) were obtained from the IRIS Data Management Center at a sampling rate of $100\,$Hz over the interval from December $2013$ to October $2015$. The raw waveforms were corrected for instrument response to obtain ground displacement and subsequently demeaned and detrended. The continuous records were divided into non-overlapping $30$-minute windows. After applying an anti-alias low-pass filter, the data were downsampled to $1\,$Hz, which is appropriate for the long-period surface waves analyzed here. The noise is thus segmented into a total of $M_{ij}\approx10000$ time windows for each station pair $(i,j)$.

Each time window was temporally normalized using a running absolute mean and spectrally whitened in the $0.009$--$0.499\,$Hz frequency band to suppress the influence of the source spectrum and stabilize the noise amplitude. The preprocessed traces were then cross-correlated for all available station pairs using FastXC~\citep{wang2025high}, with a maximum time lag of $\pm500$~s. The resulting cross-correlations $\mathbf{w}_{ij}$ were tapered and decomposed into causal and acausal branches as in Equation~3. Finally, the correlations were band-pass filtered into two period bands, $3$--$8\,$s and $8$--$50\,$s, for subsequent analysis.

\section*{Text S3. Symmetric variational autoencoder with Gumbel-Softmax trick}

\begin{figure}
\centering
\noindent\includegraphics[width=0.9\columnwidth]{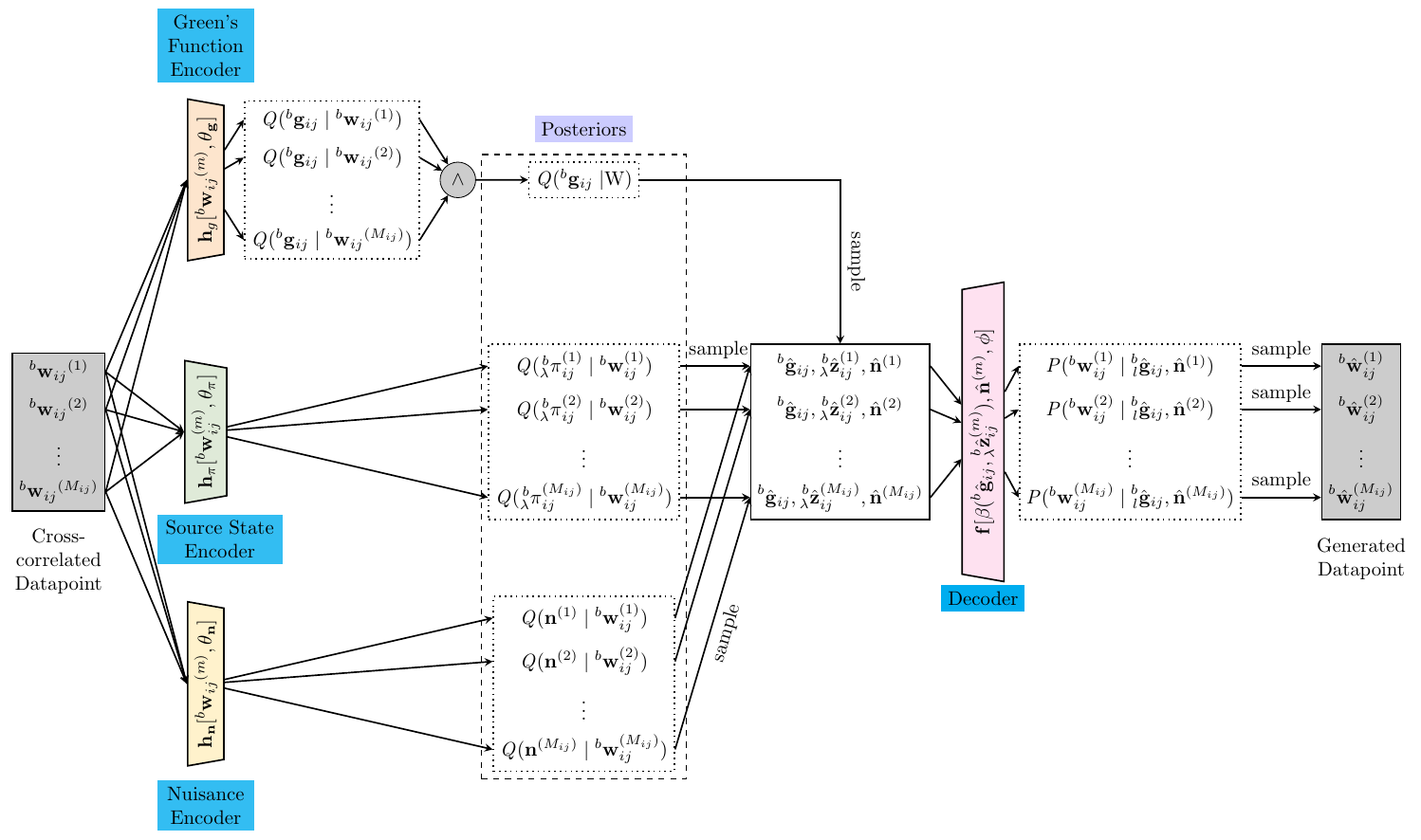}
\caption{Variational symmetric autoencoder architecture with categorical distribution. The Green's function encoder and nuisance encoder networks address the inverse problem by estimating the posterior distributions. The Green's function encoder accumulates information coherent to the cross-correlated datapoint specific to each source state. The source state encoder learns the distribution for $\lambda$ classes. The nuisance encoder encodes relevant nuisance information for each time window. The decoder generates cross-correlated windows by executing forward modeling with the latent codes as input.}
\label{fig:SymVAE}
\end{figure}

The latent-variable modeling framework introduced in Section~$2.3$ is implemented using a
symmetric variational autoencoder (SymVAE). Variational autoencoders provide a probabilistic formulation in which latent variables follow learned distributions rather
than fixed values \citep{kingma2013auto,doersch2016tutorial,bishop2023deep,prince2023understanding}.
This enables the model to jointly perform inference and generation while uncovering an interpretable low-dimensional structure in high-dimensional seismic data.
To model this discrete latent variable $\lambda$ within a differentiable framework, we adopt a categorical latent distribution implemented using Gumbel-Softmax reparameterization. 
The SymVAE architecture builds on previous work that combined symmetric autoencoders \citep{bharadwaj2022redatuming,bajad2025symmetric} with
variational modeling \citep{bharadwaj2024extracting}. Here, the encoder is designed to be permutation-invariant so that it aggregates coherent information across all windows belonging to a given station pair, while the decoder
reconstructs window embeddings symmetrically. This architectural structure ensures that
the latent state $\lambda$ correlates with physically meaningful source-state behavior,
consistent with the CSS formulation described in the main text. Figure~\ref{fig:SymVAE} illustrates the architecture of this advanced SymVAE.

To define this representation more rigorously, consider $\prescript{b}{}{\mathbf{w}}_{ij}^{(m)}$ from Equation~$1$, with $m=1,\,\ldots,\,\text{M}_{ij}$. The SymVAE framework factorizes the latent space into three physically interpretable components
\begin{itemize}
\item Green’s Function Encoder ($\mathtt{enc}_{\mathbf{g}}$). The latent variable for Green's function, $\prescript{b}{}{\mathbf{g}_{ij}}$, holds structural data that are uniform across all time windows. For each input $\prescript{b}{}{\mathbf{w}}_{ij}^{(m)}$, the encoder, described by $\theta_g$, generates a variational posterior distribution, 
    \begin{equation}
    Q(\prescript{b}{}{\mathbf{g}_{ij}}\vert \prescript{b}{}{\mathbf{w}}_{ij}^{(m)})=h_{\text{g}}[\prescript{b}{}{\mathbf{w}}_{ij}^{(m)},\theta_\text{g}],   
    \end{equation}

which serves as an approximation of the actual posterior over $\prescript{b}{}{\mathbf{g}_{ij}}$. Given that $\prescript{b}{}{\mathbf{g}_{ij}}$ reflects the shared subsurface structure, the posterior is consolidated across all data points to create a comprehensive posterior
    \begin{equation}
    Q(\prescript{b}{}{\mathbf{g}_{ij}}\vert \prescript{b}{}{\mathbf{W}}_{ij}) = Q(\prescript{b}{}{\mathbf{g}_{ij}}\vert \prescript{b}{}{\mathbf{w}}_{ij}^{(1)}) \wedge Q(\prescript{b}{}{\mathbf{g}_{ij}}\vert \prescript{b}{}{\mathbf{w}}_{ij}^{(2)}) \wedge \ldots \wedge  Q(\prescript{b}{}{\mathbf{g}_{ij}}\vert \prescript{b}{}{\mathbf{w}}_{ij}^{(M_{ij})}).
    \end{equation}
    
Here, $\wedge$ denotes the accumulation of information, and $\prescript{b}{}{\mathbf{g}_{ij}}=\mathrm{concat}\left(\prescript{b}{1}{\mathbf{g}}_{ij},\ldots,\prescript{b}{K}{\mathbf{g}}_{ij}\right)$ for $\lambda=\{1,2,\cdots,K\}$. This posterior's sampling led to a common latent vector $\prescript{b}{}{\hat{\mathbf{g}}_{ij}}$.
\end{itemize}

\begin{itemize}
    \item Nuisance Encoder ($\mathtt{enc}_{\mathbf{n}}$). The nuisance encoder identifies the incoherent variability specific to each time window. For input $\prescript{b}{}{\mathbf{w}}_{ij}^{(m)}$, the encoder, defined by $\theta_n$, generates a variational posterior given by
        \begin{equation}
            Q(\prescript{b}{}{\mathbf{n}}_{ij}^{(m)}\vert \prescript{b}{}{\mathbf{w}}_{ij}^{(m)})=h_{\text{n}}[\prescript{b}{}{\mathbf{w}}_{ij}^{(m)},\theta_\text{n}],   
        \end{equation}
        and from this a nuisance latent variable $\prescript{b}{}{\hat{\mathbf{n}}}_{ij}^{(m)}$ is drawn.
    \item Source-State Encoder ($\mathtt{enc}_{\pi}$). The source-state encoder is designed to capture time windows associated with consistent source signatures. For each $\prescript{b}{}{\mathbf{w}}_{ij}^{(m)}$, it produces a variational posterior characterized by $\theta_\pi$ as follows:
        \begin{equation}
            Q(\prescript{b}{\lambda}{\pi}_{ij}^{(m)}\vert \prescript{b}{}{\mathbf{w}}_{ij}^{(m)})=\mathrm{Gumbelsoftmax}\left(h_{\pi}[\prescript{b}{}{\mathbf{w}}_{ij}^{(m)},\theta_\pi]\right).
        \end{equation}
            To facilitate subsampling of source states, we incorporate a categorical latent distribution with class probabilities $\prescript{b}{\lambda}{\pi}_{ij}^{(m)}$. The Gumbel-softmax function serves as a differentiable proxy for categorical sampling, enabling the network to learn discrete class-based representations via backpropagation.
            Specifically, let $\lambda \in {1, \ldots,K}$ be a categorical latent variable with class probabilities $\prescript{b}{}{\pi}_{ij}^{(m)} = [\prescript{b}{1}{\pi}_{ij}^{(m)},\ldots,\prescript{b}{K}{\pi}_{ij}^{(m)}]$, sourced from the encoder as logits. To achieve a one-hot vector $\prescript{b}{\lambda}{\hat{z}}
                ^{(m)}_{ij}$ during training in a differentiable way, the following is applied:
            \begin{equation}
                \prescript{b}{\lambda}{\hat{z}}
                ^{(m)}_{ij} = \frac{\exp((\log \prescript{b}{\lambda}{\pi}_{ij}^{(m)} + g_\lambda)/\tau)}{\sum_{p=1}^K \exp((\log \prescript{b}{p}{\pi}_{ij}^{(m)} + g_p)/\tau)} \quad \text{for} \, \lambda = 1,\,\ldots,\,K,
            \end{equation}
            where $g_\lambda$ are independent and identically distributed (iid) samples of the distribution $\text{Gumbel}(0, 1)$, and $\tau$ is a parameter controlling the smoothness of the approximation. As $\tau \to 0$, the distribution becomes almost one-hot, whereas at higher $\tau$, it remains a smooth probability vector. 
            This softened categorical sample $\prescript{b}{\lambda}{\hat{z}}
            ^{(m)}_{ij}$ is inputted to the decoder during training. In inference, the class with the highest probability (i.e., $\arg\max_{\lambda} \prescript{b}{\lambda}{\pi}_{ij}^{(m)}$) is selected, or the categorical distribution is sampled directly.
            The application of the Gumbel-Softmax trick allows SymVAE to develop an organized and interpretable latent space while preserving end-to-end differentiability. This discrete representation effectively aligns with the classification of seismic events or noise patterns into unique clusters according to their waveform characteristics.
    \item Decoder: The decoder, parameterized by $\phi$, utilizes the latent triplet $[\prescript{b}{}{\hat{\mathbf{g}}_{ij}},\prescript{b}{\lambda}{\hat{z}}^{(m)}_{ij},\prescript{b}{}{\hat{\mathbf{n}}}_{ij}^{(m)}]$ to establish the conditional likelihood
    \begin{equation}
        P(\prescript{b}{}{\mathbf{w}}_{ij}^{(m)}\vert \prescript{b}{l}{\hat{\mathbf{g}}_{ij}},\prescript{b}{}{\hat{\mathbf{n}}}_{ij}^{(m)})=\text{f}[\beta(\prescript{b}{}{\hat{\mathbf{g}}_{ij}},\prescript{b}{\lambda}{\hat{z}}^{(m)}_{ij}),\prescript{b}{}{\hat{\mathbf{n}}}_{ij}^{(m)},\phi],
    \end{equation}
        where $\beta(\cdot)$ signifies the weighted sum of the structural components according to the class probabilities. The decoder samples the reconstructed waveform as
    \begin{equation}
        \prescript{b}{}{\hat{\mathbf{w}}}_{ij}^{(m)}=P\left(\prescript{b}{}{\mathbf{w}}_{ij}^{(m)}\vert\prescript{b}{l}{\hat{\mathbf{g}}_{ij}},\prescript{b}{}{\hat{\mathbf{n}}}_{ij}^{(m)}\right).
    \end{equation}
\end{itemize}
A comprehensive explanation of the formulation of SymVAEs is provided by~\cite{bharadwaj2024extracting}.
Within our framework, the loss for SymVAE is based on the evidence lower bound (ELBO) principle, but adapted to accommodate both Gaussian continuous latent variables and categorical latent variables, which are parameterized using the Gumbel-softmax. The comprehensive loss is represented as $L$ and consists of
\begin{itemize}
        \item Reconstruction term (negative log-likelihood): The decoder estimates both a mean reconstruction and a log-variance for each data point $\prescript{b}{}{\mathbf{W}}_{ij}$. Assuming a Gaussian likelihood, the formula for the negative log-likelihood is
         \begin{equation}
             L_{reconstruction}=\frac{1}{2}\ \sum_{i \not= j=1,m=1}^{n,\text{M}_{ij}}\left[\frac{{\left(\prescript{b}{}{\mathbf{w}}_{ij}^{(m)} -\prescript{b}{}{\hat{\mathbf{w}}}_{ij}^{(m)}\right)^{2}}}{\exp{\prescript{b}{}{\hat{\sigma}}_{ij}^{(m)}}}\,+\prescript{b}{}{\hat{\sigma}}_{ij}^{(m)}\right].  
         \end{equation}
    Here, $\prescript{b}{}{\hat{\sigma}}_{ij}^{(m)}$ is the logarithmic variance predicted by the network for the reconstructed waveform $\prescript{b}{}{\hat{\mathbf{w}}}_{ij}^{(m)}$. The negative log-likelihood term evaluates the fit of the data and penalizes both the squared error in the reconstruction and the uncertainty assessment.
        \item Kullback-Leibler (KL) divergence for continuous latent variables ($\prescript{b}{}{g}_{ij}$ and $\prescript{b}{}{n}_{ij}^{(m)}$): Let $\mu$ and $\text{log}\sigma$ represent the encoder's mean and log-standard deviation for a latent Gaussian variable. The KL divergence between the approximate posterior $q(\text{z} \vert \text{x})=N(\mu\vert\sigma^2)$ and a unit Gaussian prior $N(0,I)$ is given by
        \begin{equation}
        L_{KL,\,Gaussian}= \frac{1}{2}\ \sum_{i \not= j=1,m=1}^{n,\text{M}_{ij}}\left(\text{e}^{2\log\prescript{b}{}{\hat{\sigma}}_{ij}^{(m)}}+\log\prescript{b}{}{\mu}_{ij}^{(m)^2}-1-2\log\prescript{b}{}{\hat{\sigma}}_{ij}^{(m)}\right).
        \end{equation}
        This penalty is weighted by a factor $\beta_c$, which determines the strength of the Gaussian KL penalty (as in the $\beta$-VAE framework).
        \item Kullback-Leibler (KL) divergence for categorical latent variable ($\prescript{b}{\lambda}{\pi}_{ij}^{(m)}$): For categorical latent variables, we compute the posterior probabilities via softmax:
            \begin{equation}
                q(y=\lambda\vert \prescript{b}{}{\mathbf{w}}_{ij}^{(m)})=\prescript{b}{\lambda}{\pi}_{ij}^{(m)}.
            \end{equation}
            The KL divergence with respect to a categorical prior distribution $p(y)$ is
            \begin{equation}
            L_{KL,\,categorical}= \sum_{\lambda=1}^{K}\prescript{b}{\lambda}{\pi}_{ij}^{(m)}\log\frac{\prescript{b}{\lambda}{\pi}_{ij}^{(m)}}{p(y=\lambda)} .
            \end{equation}
            To ensure numerical stability, both $q(y \vert x)$ and $p(y)$ are clamped away from zero. The contribution is weighted by $\beta_\lambda$. Generally, the categorical prior is considered to be uniform.
\end{itemize}

This formulation ensures a balance between accurate reconstructions and regularization of latent distributions, enabling SymVAE to capture both continuous and discrete structures in the data.

\section*{Text S4. Eastern US dataset}

\begin{figure}
\centering
\noindent\includegraphics[width=0.8\columnwidth]{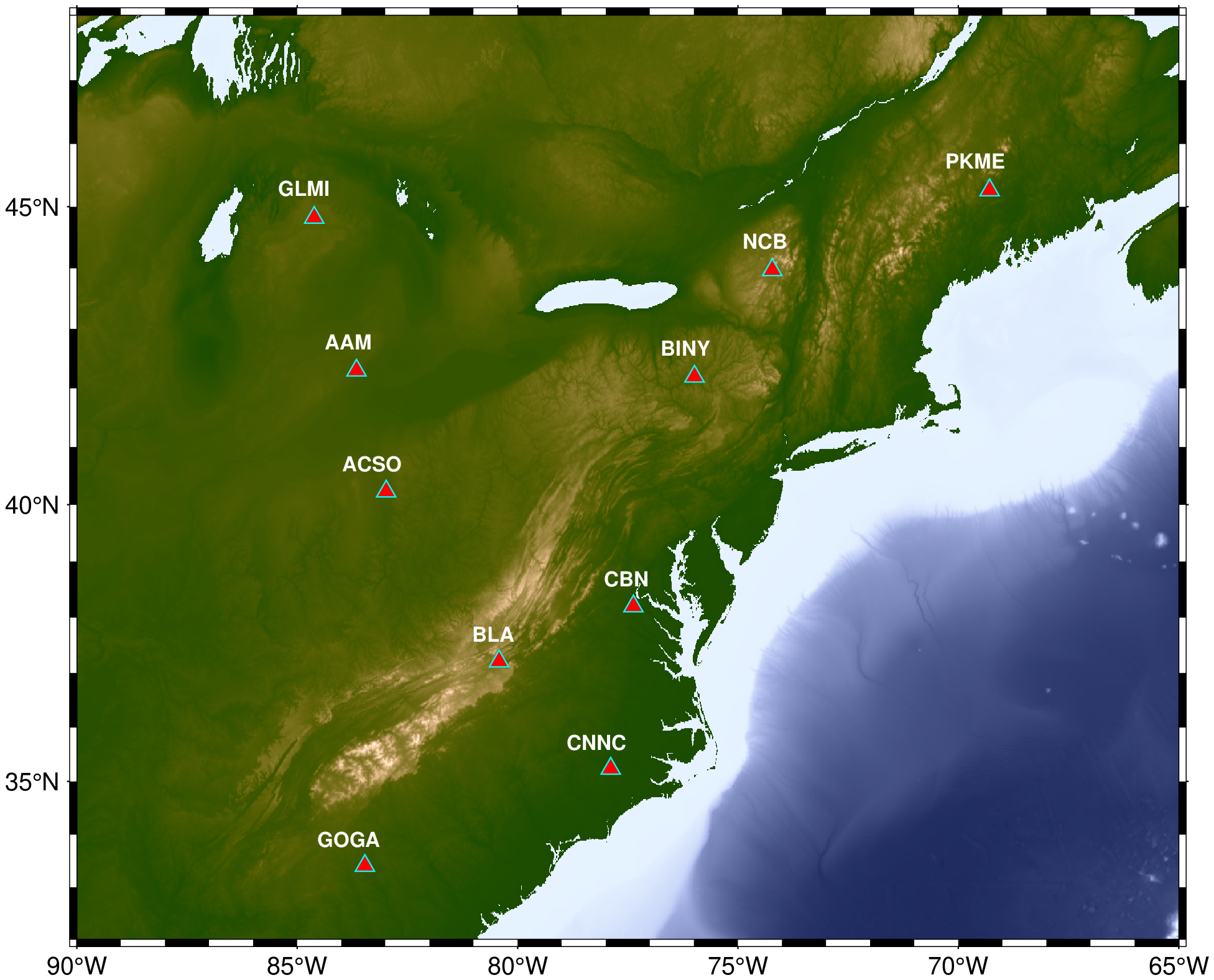}
\caption{Spatial distribution of the stations included in the eastern United States analysis.}
\label{fig:map_EUS}
\end{figure}

We also analyzed the continuous ambient seismic data from the eastern United States, a stable cratonic region characterized by low to moderate seismicity and relatively homogeneous crustal structure~\citep{liang2008ambient}. In this case, low-frequency vertical-component (LHZ) displacement data from $10$ stations of the US network (Figure~\ref{fig:map_EUS}) were recorded at a sampling rate of $1\,$Hz over the interval from January $2005$ through December $2007$ to analyze Rayleigh-wave propagation.
The waveform data were corrected for the instrument response, detrended, demeaned, and decimated to a sampling rate of $0.5\,$Hz. 
The noise is segmented into non-overlapping $30$ minute windows with a total $M_{ij}$ time windows for each pair $(i,j)$.
%
Before cross-correlation, each window was temporally normalized using a running absolute mean and was spectrally whitened in the $0.009$--$0.249\,$Hz frequency band to minimize the effects of the source signature.
The preprocessed traces were then cross-correlated for all available station pairs using FastXC~\citep{wang2025high}, with a maximum time lag of $\pm1000$~s.
%
The resulting cross-correlations $\mathbf{w}_{ij}$ were then decomposed into their causal and acausal branches as in Equation~3.  

\begin{figure}[ht]
\includegraphics[width=0.94\columnwidth]{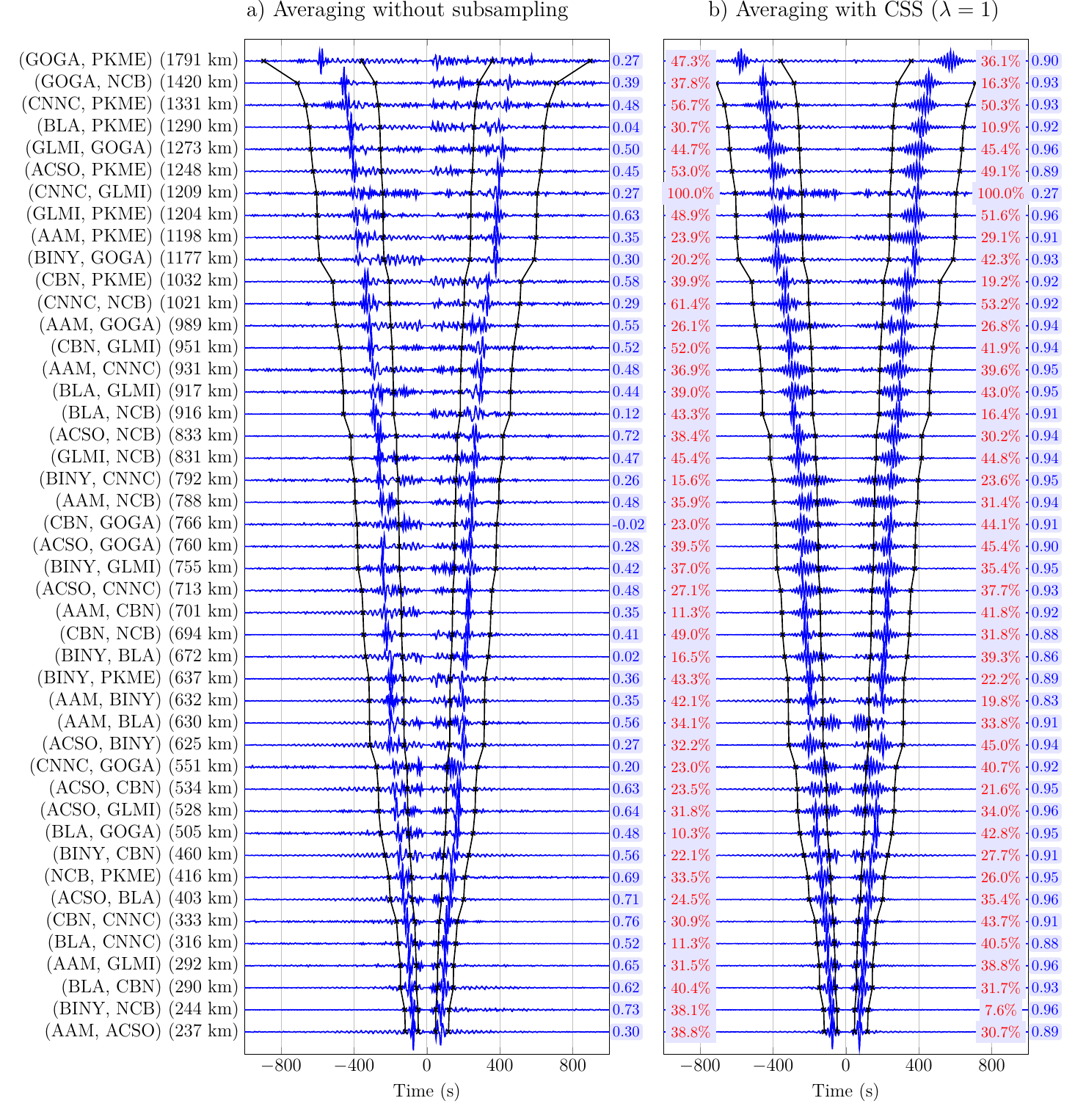}
\caption{Cross-correlation gathers for the eastern United States.  
(a) Before subsampling: The linearly averaged cross-correlations, $\widehat{\mathbb{E}}\left[ \prescript{b}{}{\mathbf{w}}_{ij} \right]$, show large amplitudes close to zero lag and pronounced causal-acausal asymmetry, indicating a directionally biased ambient noise field.  
(b) After subsampling: The conditional averages,  
$\widehat{\mathbb{E}}\left[ \prescript{b}{}{\mathbf{w}}_{ij} \mid \lambda=1 \right]$, obtained after subsampling, display enhanced waveform coherence and substantially improved causal–acausal symmetry. The percentages of time-window allocation (in red) indicate the fraction of windows selected for the source state $\lambda\text{ = 1}$. The correlation values (in blue) measure the degree of symmetry between the acausal and causal components for each station pair. The black lines mark the time lags corresponding to velocities of $2$ and $5\,$km/s.}
\label{fig:gathers_EUS}
\end{figure}

\begin{figure}[ht]
\centering
\includegraphics[width=0.75\columnwidth]
{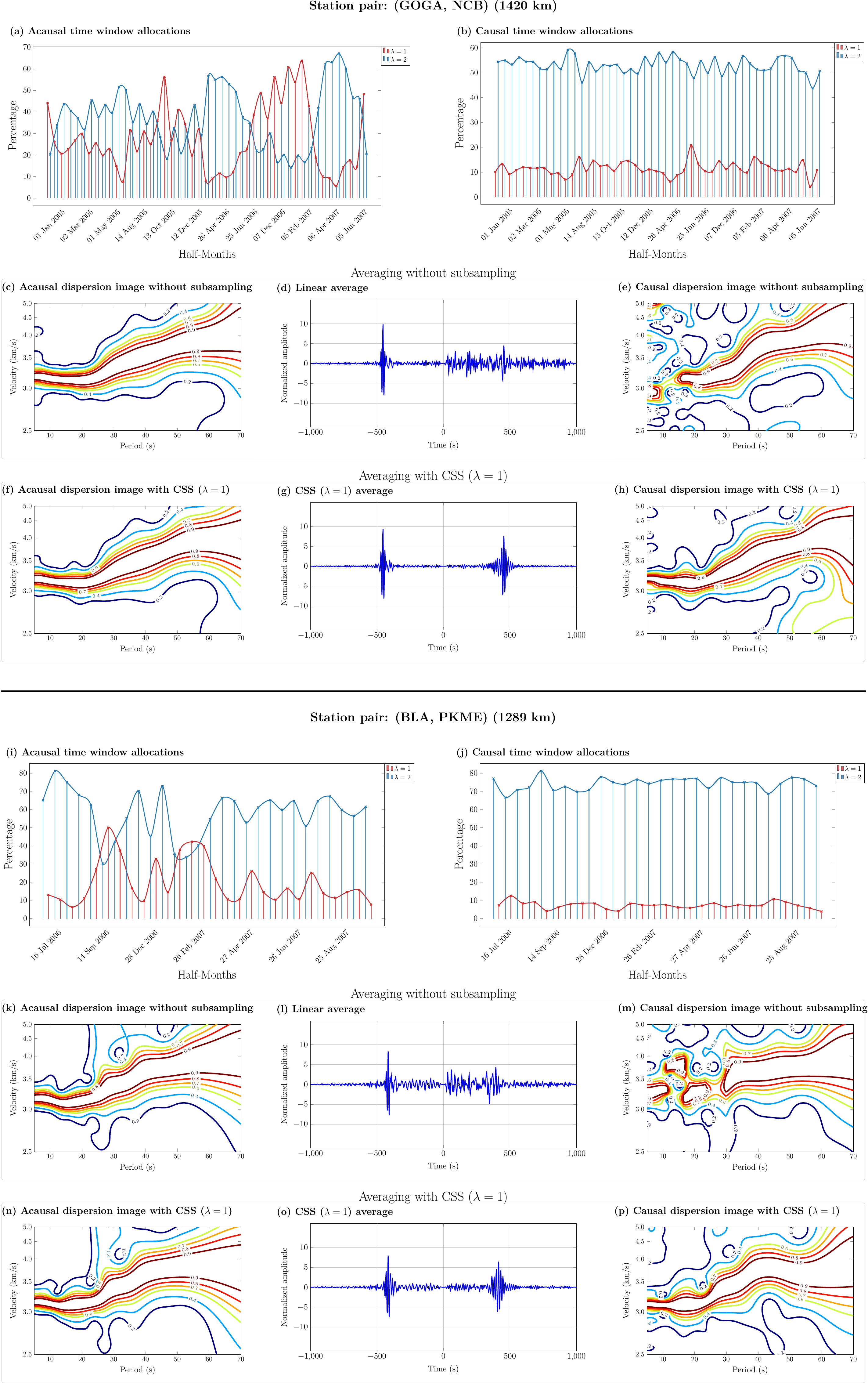}
\caption{
Temporal and dispersion analysis for the station pairs (GOGA, NCB) and (BLA, PKME).  
(a–b, i–j) Acausal and causal time-window selections obtained from CSS for source states $\lambda\text{ = \{1, 2\} }$.  
(c–e, k–m) Before subsampling: standard linear averaging produces dispersion images and cross-correlations with low coherence and pronounced causal–acausal asymmetry.  
(f-h, n-p) After subsampling: source state $\lambda\text{ = 1}$ results in dispersion images and cross-correlations with substantially improved causal–acausal symmetry.
}
\label{fig:assign_and_symm_EUS}
\end{figure}

Figure~\ref{fig:gathers_EUS} compares linearly averaged cross-correlations, $\widehat{\mathbb{E}}\left[ \prescript{b}{}{\mathbf{w}}_{ij} \right]$, with conditionally averaged cross-correlations, $\widehat{\mathbb{E}}\left[ \prescript{b}{}{\mathbf{w}}_{ij} \mid \lambda=1 \right]$, for several station pairs. 
The linear averages (Figure~\ref{fig:gathers_EUS}a) display limited causal–acausal symmetry, reflecting the non-uniform distribution of ambient noise sources.
In contrast, the conditionally averaged correlations (subsampled averages in Figure~\ref{fig:gathers_EUS}b) exhibit a markedly stronger symmetry: the Rayleigh waves on the causal and acausal branches are nearly identical. This enhanced symmetry is quantified in Figure~\ref{fig:gathers_EUS} and the figure also reports, for each pair of stations, the fraction $\frac{\left|\prescript{b}{\lambda}{\mathcal{C}}_{ij}\right|}{M_{ij}}$ of time windows retained in the causal and acausal branches by CSS. 
These fractions indicate that nearly symmetric cross-correlations can be obtained from only a modest subset of the available windows, typically about $10$–$50\%$ of the total.
Further examination of the retained time windows, illustrated in Figures~\ref{fig:assign_and_symm_EUS}a--b and \ref{fig:assign_and_symm_EUS}i--j, reveals that the causal and acausal branches generally select different, non-random subsets of time windows.
For both pairs (GOGA, NCB) and (BLA, PKME) in Figure~\ref{fig:assign_and_symm_EUS}, only a small fraction (nearly $10\%$) of windows in each half-month contribute to the state $\lambda = 1$ for the causal branch.
This indicates that the strongest ambient noise sources are probably located outside the stationary zone, while the stationary zone itself is dominated by weaker and less frequent sources.
On the other hand, the acausal branch selects a larger fraction of windows (up to $50\%$) in most half-month intervals, suggesting a more consistent presence of coherent sources within the stationary zone for that branch --- therefore, the linear average in this case already exhibits a higher signal-to-noise ratio.



The improvement in cross-correlation symmetry translates directly into enhanced Rayleigh-wave dispersion measurements.
Figures~\ref{fig:assign_and_symm_EUS}c--e for (GOGA, NCB) and \ref{fig:assign_and_symm_EUS}k--m for (BLA, PKME) display the Rayleigh-wave dispersion images obtained without subsampling, which show asymmetry between the causal and acausal branches. In practice, such an asymmetry has often led studies to rely preferentially on the higher-SNR branch when extracting dispersion information. However, selecting only the “better” side does not guarantee that the retained energy predominantly originates from the stationary phase zone or that it represents the physical interstation response.
After applying CSS averaging for $\lambda = 1$, the dispersion images (Figures \ref{fig:assign_and_symm_EUS}f--h and \ref{fig:assign_and_symm_EUS}n--p) exhibit markedly enhanced coherence and symmetry. This improved balance between the causal and acausal branches enables a much more reliable picking of group velocities. 

\begin{figure}[ht]
\centering
\includegraphics[width=\columnwidth]{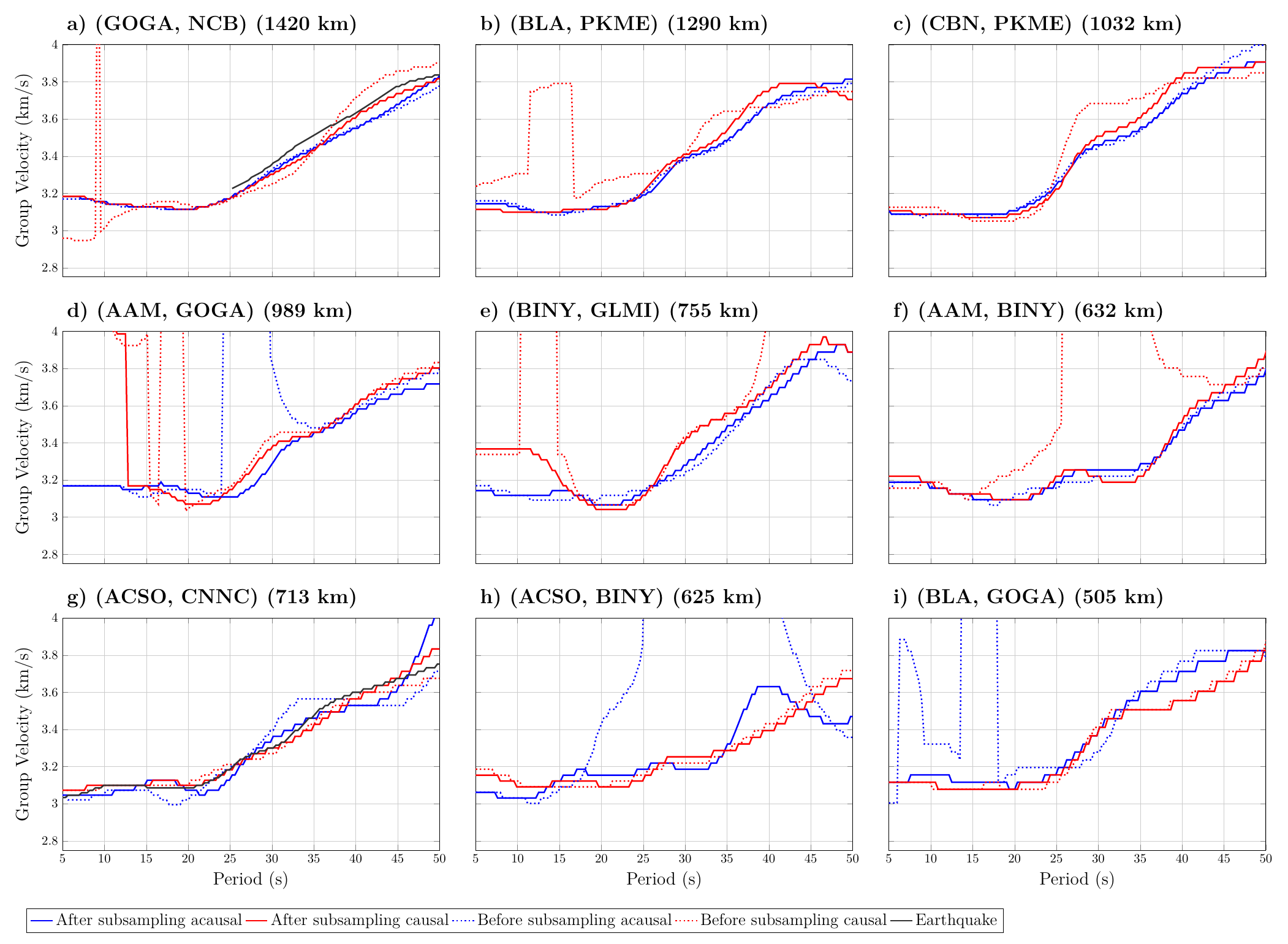}
\caption{Group‐velocity dispersion curves for nine station pairs in the eastern United States. Solid lines represent the causal and acausal dispersion curves obtained for source state $\lambda\text{ = 1}$, while dotted lines indicate the corresponding dispersion curves based on linear averaging. Earthquake‐based dispersion curves are shown for comparison for station pairs (GOGA, NCB) and (ACSO, CNNC). For all station pairs, the dispersion curves retrieved using source state $\lambda\text{=1}$ show markedly reduced scatter between $10\,$ and $50\,$s, whereas the linearly averaged curves display substantial variability and systematically underestimate velocities. This closer agreement demonstrates the effectiveness of CSS for recovering symmetric dispersion information.
}
\label{fig:dispersion_3x3}
\end{figure}

Figures~\ref{fig:dispersion_3x3}a--i show group-velocity dispersion curves, extracted by automated picking, for nine station pairs with short, intermediate, and long interstation distances.
Again, for each pair, the dispersion curves obtained by simple linear averaging display substantial scatter and a clear mismatch between the causal and acausal branches. After applying conditional averaging, the dispersion estimates become much more stable: the causal and acausal branches closely coincide, producing nearly symmetric behavior in the $5–50\,$s period band, where Rayleigh-wave energy is strongest. The pairs with the longest and intermediate distances, including (GOGA, NCB) ($1420\,$km), (BLA, PKME) ($1290\,$km), (CBN, PKME) ($1032\,$km), and (ACSO, CNNC) ($713\,$km), show the most pronounced improvements, with the subsampled curves collapsing onto a well-defined trend. 
 Other station pairs, such as (AAM, GOGA) ($989\,$km), (BINY, GLMI) ($755\,$km), (AAM, BINY) ($632\,$km), (ACSO, BINY) ($625\,$km), and (BLA, GOGA) ($505\,$km), also exhibit notably closer agreement between the causal and acausal branches than in the linear averages.
Finally, we compare the CSS-derived dispersion curves with the earthquake-derived dispersion curves for (GOGA, NCB) and (ACSO, CNNC) pairs, using earthquakes of about $M_w\,4.5$ located near the GOGA and ACSO stations, respectively. For both pairs, the subsampled symmetric dispersion curves agree well with the earthquake-based dispersion curves.

\begin{figure}
\centering
\noindent\includegraphics[height=0.975\columnwidth]{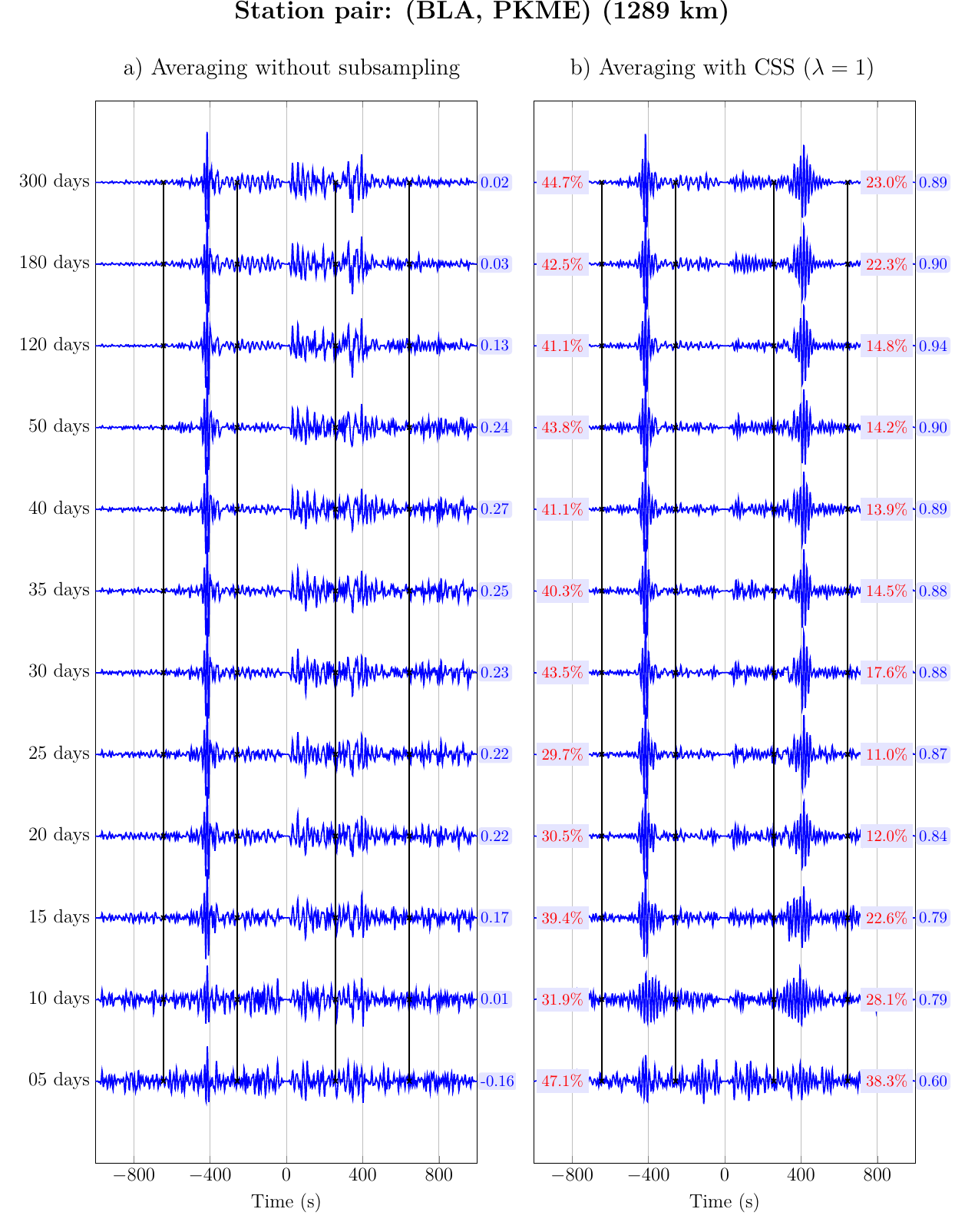}
\caption{Comparison of averaged ambient-noise cross-correlations for the station pair (BLA, PKME) for increasing durations ($5–300$ days): (a) averaging without subsampling and (b) averaging with CSS ($\lambda\text{=1}$). Linear averaging without subsampling shows pronounced causal–acausal asymmetry and low signal to noise ratio, particularly at shorter averaging durations. In contrast, CSS produces more symmetric and stable correlations even for substantially shorter durations. Red percentages denote the fraction of data retained after subsampling, and blue values quantify the degree of causal–acausal symmetry.}
\label{fig:Monthly_comparison}
\end{figure}

CSS is particularly effective for short-duration surveys, where linear averaging fails to exhibit high signal-to-noise ratios for both branches.
To demonstrate this, we analyzed the station pair (BLA, PKME) by training on progressively longer short‑duration subsets of the data—rather than on the full two‑year record. The results are presented in Figure~\ref{fig:Monthly_comparison}, where, in contrast to linear averaging, CSS attains clear causal--acausal symmetry even with short records (as few as $10$ days), and the symmetry strengthens as the data duration increases.
%

As shown in Figure~\ref{fig:assign_and_symm_EUS}, 
the subsampling of neither pair exhibits pronounced seasonal modulation over the two-year interval.
This lack of seasonal patterns in the selected windows highlights an important distinction: CSS conditioning is driven by the learned representation of stationary-phase contributions rather than by external seasonal variables (e.g., time of year or ocean microseism indices), demonstrating that the methods using predefined temporal or environmental markers are not sufficient for effective subsampling.

\bibliographystyle{plainnat}
\bibliography{references_supplementary}